\documentclass[aps,prb,amsmath,amssymb,amsfonts,preprintnumbers,groupedaddress]{revtex4}
\usepackage{graphicx}
\graphicspath{{./}}
\usepackage[caption=false]{subfig}
\usepackage{makecell}
\usepackage{hhline}
\usepackage{tabularx}
\newcolumntype{Y}{>{\centering\arraybackslash}X}

\usepackage{booktabs,array,multirow}
\usepackage[dvipsnames]{xcolor}
\usepackage{hyperref}
\usepackage[papersize={8.5in,11in},total={7in,9.5in}]{geometry}
\usepackage[section]{placeins}
\usepackage[outdir=./]{epstopdf}
\begin{document}

\title{Optical Properties of Anisotropic Excitons in Phosphorene}
%\author{Matthew Brunetti}
%\affiliation{City Tech}
%\author{Oleg Berman}
%\affiliation{City Tech, Graduate Center}
%\author{Roman Kezerashvili}
%\affiliation{City Tech, Graduate Center}
\author{Matthew N. Brunetti$^{1,2}$, Oleg L. Berman$^{1,2}$, and Roman Ya. Kezerashvili$^{1,2}$}
\affiliation{%
$^{1}$Physics Department,  New York City College of Technology\\
The City University of New York,
  300 Jay Street,   Brooklyn NY, 11201, USA \\
$^{2}$The Graduate School and University Center\\
The City University of New York,
New York, NY 10016, USA \\
}

\date{\today}

\begin{abstract}
  We study the eigenenergies and optical properties of both direct excitons in a phosphorene monolayer in different dielectric environments, and indirect excitons in heterostructures of phosphorene with hexagonal boron nitride.
  For these systems, we solve the 2D Schr\"{o}dinger equation using the Rytova-Keldysh (RK) potential for direct, and both the RK and Coulomb potentials for indirect excitons.
  The results show that excitons formed from charge carriers with anisotropic effective mass exhibit enhanced (suppressed) optical absorption, compared to their 2D isotropic counterparts, under linearly polarized excitations along the crystal axis with relatively smaller (larger) effective carrier masses.
  This anisotropy leads to dramatically different excited states than the isotropic exciton.
  The direct exciton binding energy depends strongly on the dielectric environment, and shows good agreement with previously published data.
  For indirect excitons, the oscillator strength and absorption coefficient increase as the interlayer separation increases.
  The choice of RK or Coulomb potential does not significantly change the indirect exciton optical properties, but leads to significant differences in the binding energy for small interlayer separation.
  % Our calculations contribute to ongoing efforts to characterize the optical properties of excitons in phosphorene and invite further study of these systems.
\end{abstract}

\maketitle

\section{\label{sec:intro}Introduction}

  The experimental discovery of graphene in 2004~\cite{novoselov2004electric} was a fascinating first glimpse into the world of two-dimensional (2D) materials \textendash{} its exceptional mechanical, thermal, and electrical properties suggested a new paradigm of flexible, durable, and highly efficient 2D electronic devices.
  In 2010, when monolayers of both insulating hexagonal boron nitride ($h$-BN)~\cite{Dean2010} and semiconducting transition metal dichalcogenides (TMDCs)~\cite{Mak2010} were first exfoliated, research efforts towards the development of next-generation 2D devices accelerated.
  The discovery of monolayer (ML) black phosphorus, referred to as phosphorene, came a full decade after the advent of graphene when a flurry of publications in 2014 heralded the arrival of a new addition to the 2D materials universe~\cite{Li2014b,Koenig2014,Liu2014a,Castellanos-Gomez2014,Buscema2014,Xia2014a}.
  Within the first year of its discovery, the number of publications on phosphorene grew tenfold~\cite{Castellanos-Gomez2015}, an unprecedented rate of growth even within the rapidly expanding field of 2D materials research.
  The sudden shift in intense research focus towards phosphorene is clearly justified, due to phosphorene's unique properties which make it a promising candidate for a variety of applications unsuited to its 2D relatives.

  Perhaps the primary distinguishing feature of phosphorene is its highly corrugated crystal structure, where a single monolayer appears to be composed of two distinct planes of phosphorus atoms.
  Each atom bonds to three neighbors, two of which are in the same plane and one which occupies the opposite plane~\cite{Sorkin2017}.
  The in-plane and out-of-plane bonds are characterized by very different bond lengths and bond angles, which in turn leads to extreme anisotropy in its mechanical, thermal, and electronic properties~\cite{Fei2014,Rodin2014,Ong2014,Qin2014,Wei2014,Xu2015a,Jain2015,Chaves2015,Appalakondaiah2012,Dai2017}.
  This intrinsic structural anisotropy is expressed in nearly every property of phosphorene, manifesting itself in charge carrier effective masses~\cite{Li2014a} and mobilities~\cite{Li2014b,Liu2014a,Xia2014a,Lu2015a}, DC conductivity~\cite{Xia2014}, Raman spectra~\cite{Tran2014,Wang2015b,Hong2014a,Yuan2015a,Lu2015a,Ribeiro2015,Caklr2015,Low2014}, optical absorption~\cite{Xia2014a} and photoluminescence~\cite{Wang2015b} spectra, and in its response to mechanical strain~\cite{Wei2014}.
  Furthermore, the rippled structure of phosphorene leads to a larger surface area, which makes it ideal~\cite{Cho2016} for a variety of environmental~\cite{Mayorga-Martinez2015,Guo2016,Gu2017} and biomedical~\cite{Yew2017,Qiu2018} sensing applications.
  In addition to the already unique and intriguing structure of black phosphorus, it was recently shown that there are four more allotropes of ML phosphorus, each exhibiting unique crystal structures and distinct material properties~\cite{Wu2015a}.

  Unlike the TMDCs, which are indirect gap semiconductors for all but their ML forms~\cite{Chhowalla2013}, phosphorene remains a direct gap semiconductor from its bulk form ($\approx$ 0.3 eV)~\cite{Asahina1984} down to a single ML ($\approx$ 2 eV)~\cite{Rudenko2014}.
  In addition to its dependence on layer number~\cite{Qiao2014,Tran2014,Wang2015b,Liang2014,Liu2015a,Chen2015,Low2014a,Woomer2015}, the band gap, as well as many other properties, is also sensitive to the magnitude and direction of an applied mechanical strain~\cite{Lv2014,Ong2014,Fei2014,Li2014f,Rodin2014a,Elahi2015,Cakir2014b}, giving researchers a variety of ways to tailor the electronic properties of phosphorene to suit a particular task.
  Due to its broadly tunable band gap, phosphorene has also been identified as a promising material for converting solar energy to chemical energy~\cite{Hu2016}.

  % Since the band gaps of TMDC monolayers fall between 1 and 2 eV~\cite{Wang2012}, while in bilayer graphene an electric field can be used to create a band gap of at most 0.2 eV~\cite{Castro2007}, few-layer phosphorene's coverage of the sub-eV range is crucial for a variety of applications including fiber optic telecommunications~\cite{Soole1991}, thermal imaging, and thermoelectric applications~\cite{Zhang2014b,Lv2014,Lv2014b,Qin2014,Flores2015,Fei2014a}.
  % In addition to its applications for which neither graphene or TMDCs can be used, phosphorene also exhibits an ideal combination of material parameters that make it more well suited to traditional electronic device applications, such as certain use-cases of field-effect transistors (FETs), than either graphene or the TMDCs~\cite{Castellanos-Gomez2015}.
  % To this end, there has been considerable research interest in phosphorene as the active material in FETs~\cite{Li2014b,Buscema2014a,Deng2014,Na2014,Avsar2015,Kamalakar2015,Wang2014b,Doganov2015}, logic elements in circuits~\cite{Liu2014a,Du2014,Zhu2015d}, and photodetectors~\cite{Youngblood2015,Hong2014a,Low2014b,Buscema2015,Engel2014,Yuan2015a,Xia2014}.

  Characterizing the optical properties of phosphorene is not only important in the context of phosphorene's potential applications to optoelectronic devices, but is an essential tool in understanding its fundamental properties, e.g.\ its electronic band structure.
  Indeed, some of the first experimental studies of phosphorene measured its photoluminescence (PL)~\cite{Liu2014a,Zhang2014}, optical absorption~\cite{Xia2014a}, and Raman~\cite{Castellanos-Gomez2014} spectra, as well as photocurrent generation~\cite{Hong2014a}.
  In particular, one group observed an ``extraordinary'' PL peak in a phosphorene bilayer sample~\cite{Zhang2014}, and another found a similarly strong PL signal in ML phosphorene, centered at 1.45 eV~\cite{Liu2014a}.
  It was quickly realized that these remarkable optical properties were due to strongly bound and optically active excitons within the phosphorene ML.

  % An exciton the bound state of a conduction band electron and a valence band hole.
  % They are formed in semiconductors when the electron absorbs just enough energy (typically from a properly-tuned pump laser) to exit the valence band, leaving behind a positively charged hole.
  % The electron and hole form a bound state \textendash{} the exciton \textendash{} as a result of their mutual electrostatic attraction.
  % Excitons act like hydrogen atoms embedded within the crystal structure of the semiconductor, and their interactions with light tend to dominate the optical spectrum of the material at energies below the band gap.
  Exciton binding energies in bulk semiconductors tend to be on the order of a few tens of meV due to strong dielectric screening, adversely affecting their stability at room temperature and restricting the energy ranges in which they are optically active.
  By contrast, excitons in 2D semiconductors exhibit dramatically increased binding energies compared to their bulk counterparts, due on the one hand to quantum confinement effects reducing the degrees of freedom and therefore the average kinetic energy of the system~\cite{Kezerashvili2019}, while on the other hand experiencing stronger electrostatic screening.

  Around the same time that experimentalists first observed evidence of strongly bound and optically active excitons in mono- and few-layer phosphorene, a number of \textit{ab-initio} studies of the electronic band structure and excitonic properties of phosphorene were published.
  Using a variety of theoretical approaches and numerical methods, the binding energy of the direct exciton in a phosphorene ML was calculated to be between 0.7-0.8 eV~\cite{Tran2014,Caklr2015,Choi2015,Rodin2014,Cakir2014b,Seixas2015,Prada2015}, while the exciton binding energy of ML phosphorene on an Si/SiO$_2$ substrate was calcuated~\cite{Rodin2014} and measured~\cite{Yang2015a} to be around 0.3 eV.
  Another \textit{ab-initio} study~\cite{Tran2014} predicted that the exciton-forming optical transition centered at 1.45 eV in a freestanding phosphorene monolayer could absorb a staggering 15\% of incident light, but only if the excitation was linearly polarized along the \textit{armchair} crystal direction \textendash{} the extreme anisotropy of phosphorene left the transition almost completely dark for light polarized perpendicular to the armchair crystal axis.

  While research on excitons in phosphorene has largely focused on direct excitons in ML phosphorene, the field has recently expanded to consider spatially separated excitons formed in heterostructures (HS) consisting of two phosphorene monolayers separated by few-layer insulating $h$-BN, abbreviated as PHP HS.
  % Indirect excitons are typically made by first creating a direct exciton in a single monolayer via optical excitation, then applying a perpendicular electric field to separate the electron and hole into separate monolayers.
  In this configuration, the electron and hole occupy different parallel phosphorene monolayers and their recombination is suppressed by the tunneling barrier created by the dielectric separating the phosphorene.
  As a result, indirect excitons exhibit much longer lifetimes than their direct counterparts.
  Indirect excitons (also called dipolar excitons due to their intrinsic dipole moment) exhibit many of the same properties as direct excitons, but importantly, this intrinsic dipole moment creates a weak, repulsive exciton-exciton interaction.
  As a result of their enhanced binding energy, the typically small effective mass of charge carriers in semiconductors, and weakly repulsive inter-particle interactions, indirect excitons in e.g.\ the TMDCs have been identified by theorists as promising candidates for high-temperature Bose-Einstein condensation (BEC) and superfluidity~\cite{Fogler2014,Berman2016a,Berman2017a}.
  By extension, indirect excitons in phosphorene recently attracted interest when it was proposed that they could exhibit directionally-dependent BEC and superfluidity~\cite{Berman2017,Saberi-Pouya2018}.
  % In particular, the critical velocity of superfluidity, the spectrum of collective excitation, superfluid and normal component concentrations, and mean-field critical temperature of superfluidity were all found to be anisotropic.

  In this work, we calculate the eigenfunctions and eigenenergies of (i) direct excitons in ML phosphorene and (ii) indirect excitons in a PHP HS.
  % In this work, we calculate the optical transition energies, oscillator strengths and absorption coefficients for optical transitions from the excitonic ground state to the excitonic excited states in three different scenarios: i) electrons and holes occupying the same phosphorene monolayer (direct excitons), and spatially separated electrons and holes occupying parallel phosphorene monolayers (indirect excitons) which are ii) stacked directly on top of each other (phosphorene bilayer), and iii) separated by one or more monolayers of $h$-BN (phosphorene/$h$-BN heterostructure).
  First, the Schr\"{o}dinger equation for an interacting electron and hole with anisotropic effective masses is solved numerically, yielding the eigenenergies and corresponding eigenfunctions of the excitonic ground and excited states.
  We then use well-established methods~\cite{Snokebook,Lozovik1997,Brunetti2018a,Brunetti2018b} for analyzing inter-excitonic optical transitions to study the anisotropic exciton eigensystem, obtaining the optical transition energies, oscillator strengths and absorption coefficients of the transitions.

  This paper is organized as follows.
  In Sec.~\ref{sec:theory}, we summarize the theoretical approach for solving the 2D Schr\"{o}dinger equation of the electron-hole system with anisotropic effective masses.
  In Sec.~\ref{sec:optics}, we present the theoretical framework for calculating the optical properties of excitons.
  We describe our computational approach and discuss the choice of input parameters in Sec.~\ref{sec:approach}.
  The results of our calculations for direct and indirect excitons follow in Sec.~\ref{sec:results}.
  We compare the calculated properties of excitons in phosphorene to the properties of excitons in other 2D materials in Sec.~\ref{sec:analysis}.
  Our conclusions follow in Sec.~\ref{sec:conclusion}.

\section{\label{sec:theory}Excitons with anisotropic effective mass}

  In order to analyze the optical properties of excitons in phosphorene, we must first calculate the eigenenergies and eigenfunctions of the exciton by solving the Schr\"{o}dinger equation describing an interacting electron and hole with anisotropic effective masses.
  This in turn requires providing the material properties of phosphorene as input parameters to the Schr\"{o}dinger equation, in particular the anisotropic effective carrier masses, the dielectric screening length of ML phosphorene, and the dielectric constant of the environment.
  Significant effort has already been dedicated to characterizing the electronic structure of phosphorene both experimentally~\cite{Liang2014,Zhang2014} and using theoretical~\cite{Li2014a} and \textit{ab-initio} techniques~\cite{Du2010,Qiao2014,Dai2017,Li2014f,Caklr2015,Rudenko2014,Xu2015a}.
  Importantly, these analyses have yielded, among other things, the anisotropic effective masses of both electrons and holes.
  Both the static dielectric constant~\cite{Asahina1984} and ML thickness of phosphorene~\cite{Kumar2016} are also known, which is important for characterizing the electrostatic interaction between the electron and hole.
  These parameters can be inserted directly into the Schr\"{o}dinger equation describing the electron-hole system, enabling a straightforward solution of the anisotropic exciton eigensystem.
  We therefore present the quantum mechanical description of the electron and hole using the 2D Schr\"{o}dinger equation in such a way that it can be applied to either direct or indirect excitons \textendash{} further discussion on the formal differences between the two systems will be given as necessary.
  Finally, we note that the orthogonal crystal axes of phosphorene are referred to as the \textit{armchair} and \textit{zigzag} directions~\cite{Wei2014,Guo2014,Li2014c,Li2018,Jain2015,Zhang2014b,Peng2014} \textendash{} following the convention in the literature~\cite{Tran2014,Caklr2015,Rodin2014}, we associate the $x$- and $y$-axes with the \textit{armchair} and \textit{zigzag} directions, respectively.

  % Applying the effective mass approximation for electrons (holes) in the conduction (valence) bands of phosphorene, the Hamiltonian describing the interaction between an isolated electron-hole pair with anisotropic effective mass, where the electron and hole are bound to their respective phosphorene monolayers, is given by:
  Within the effective mass approximation, the Hamiltonian for an interacting electron and hole with anisotropic effective mass, constrained to move in the plane of their respective monolayers, is given by:

  \begin{equation}
	\hat{H}_0 = \frac{- \hbar^2}{2} \left( \frac{1}{m_e^x} \frac{\partial^2}{\partial x_e^2} + \frac{1}{m_e^y} \frac{\partial^2}{\partial y_e^2} + \frac{1}{m_h^x} \frac{\partial^2}{\partial x_h^2} + \frac{1}{m_h^y} \frac{\partial^2}{\partial y_h^2} \right) + V\left( \mathbf{r}_e - \mathbf{r}_h \right),
	\label{eq:hamehxy}
  \end{equation}
  where the $m_i^j,~j=x,y,~i=e,h$ correspond to the effective mass of the electron or hole in the $x$ or $y$ direction, respectively, the positions of the electron and hole are given by $\mathbf{r}_i = (x_i,y_i,z_i)$, and $V\left( \mathbf{r}_e - \mathbf{r_h} \right)$ describes the electrostatic interaction between the electron and hole.
  Eq.~\eqref{eq:hamehxy} can be used to treat both direct excitons ($z_e - z_h = 0$) and indirect excitons ($z_e - z_h \equiv D$), where the interlayer separation $D = l_{phos} + N_{\text{BN}} l_{\text{BN}}$ is the distance between the middle of the phosphorene monolayers, $l_{phos}$ and $l_{\text{BN}}$ are the thicknesses of ML phosphorene and $h$-BN, respectively, and $N_{\text{BN}}$ is the number of $h$-BN monolayers separating the phosphorene.
  For indirect excitons in a PHP HS, we consider the average $z$-position of the electron and hole to be in the middle of their respective phosphorene monolayers, since the narrow vertical confinement of each particle resembles a one-dimensional particle in a box.
  Therefore, the electron-hole separation $D$ must account for (i) the thickness of one phosphorene ML, which can be pictured as being ``split'' between the upper half of the lower ML and the lower half of the upper ML, as well as (ii) the vertical distance between the phosphorene monolayers themselves due to the intervening $h$-BN.
  %   %   For 2D materials in general, the distinction between the physical thickness of a monolayer and the interlayer spacing between two or more stacked monolayers is often unclear.
  %   For example, an \textit{ab-initio} study~\cite{Kumar2016} of phosphorene determined that the upper and lower sub-planes of a single phosphorene monolayer are separated by a distance of 0.221 nm (the monolayer thickness), while the distance between the top of one phosphorene monolayer to the bottom of another phosphorene monolayer was determined to be 0.32 nm (the interlayer separation).
  %   To further complicate matters, atomic force microscopy (AFM) measurements of the ``thickness'' of a phosphorene monolayer have been reported to be between 0.7 nm~\cite{Wang2015b} and 0.85 nm~\cite{Liu2014a}, though the authors of both papers acknowledge that AFM measurements of monolayer thickness ``tend to exceed theoretical results''~\cite{Liu2014a}, and that the ``nominal thickness''~\cite{Wang2015b} of a phosphorene monolayer is instead about 0.53 nm, which agrees well with the value of $0.221~\text{nm} + 0.32~\text{nm}=0.541~\text{nm}$ which can be derived from Ref.~\onlinecite{Kumar2016}.
  % Ultimately, from the perspective of this study, the distinction between ``monolayer thickness'' and ``interlayer spacing'' is moot.

  Applying the standard procedure for separation of variables in the two-body problem~\cite{LandauLif} to the anisotropic Hamiltonian~\eqref{eq:hamehxy}, we define the center-of-mass coordinate as $\mathbf{R} = (X,Y)$, $X = (m_e^x x_e + m_h^x x_h)/(m_e^x + m_h^x)$, $Y = (m_e^y y_e + m_h^y y_h)/(m_e^y + m_h^y)$, and the relative separation between the electron and hole as $\mathbf{r} = \mathbf{r}_e - \mathbf{r}_h = (x,y,D)$, $x = x_e - x_h$, $y = y_e - y_h$, $D = z_e - z_h$.
  After separation of variables in~\eqref{eq:hamehxy}, the Schr\"{o}dinger equation for the relative motion of the electron-hole system is given by:

  \begin{equation}
	\left[ -\frac{\hbar^2}{2 \mu^x} \frac{\partial^2}{\partial x^2} - \frac{\hbar^2}{2 \mu^y} \frac{\partial^2}{\partial y^2} + V\left( \mathbf{r} \right) \right] \psi\left( \mathbf{r} \right) = E \psi\left( \mathbf{r} \right),
	\label{eq:relschro}
  \end{equation}
  where $\mu^{j} = (m_e^{j} m_h^{j})/(m_e^{j} + m_h^{j})$, $j=x,y$, is the reduced mass of the exciton in the $x$ and $y$ directions and $E$ and $\psi\left( \mathbf{r} \right)$ are the eigenenergies and eigenfunctions of the exciton, respectively.

  For the direct exciton, the electron-hole interaction $V\left( \mathbf{r} \right)$ is described by the Rytova-Keldysh (RK) potential~\cite{Rytova1967,Keldysh1979},

  \begin{equation}
	V_{\text{RK}} \left( \mathbf{r} \right) \equiv V_{\text{RK}} \left( r \right) = -\frac{\pi k e^2}{2 \kappa \rho_0} \left[ H_0 \left( \frac{r}{\rho_0} \right) - Y_0 \left( \frac{r}{\rho_0} \right) \right].
	\label{eq:vkeld}
  \end{equation}
  In Eq.~\eqref{eq:vkeld}, $k = 9 \times 10^9$ N$\cdot$m$^2$/C$^2$, $r \equiv \lvert \mathbf{r} \rvert = \sqrt{x^2 + y^2}$ (for the direct exciton in a phosphorene ML) or $r = \sqrt{x^2 + y^2 + D^2}$ (for the indirect exciton in a PHP HS) is the magnitude of the relative electron-hole separation, $\kappa = (\epsilon_1 + \epsilon_2)/2$ describes the surrounding dielectric environment, where $\epsilon_1$ and $\epsilon_2$ correspond to the dielectric constants of the materials a) above and below the ML for the direct exciton, or b) between and surrounding the phosphorene monolayers for the indirect exciton in a PHP HS, $H_0$ and $Y_0$ are the Struve and Bessel functions of the second kind, respectively, and $\rho_0$ is the screening length, given by~\cite{Cudazzo2011,Berkelbach2013}:

  \begin{equation}
	\rho_0 = \frac{2 \pi \chi_{2D}}{\kappa},
	\label{eq:rho0chi2d}
  \end{equation}
  where $\chi_{2D}$ is the 2D polarizability, which can be calculated via \textit{ab-initio} methods.

  The asymptotic behavior of the RK potential with respect to the interparticle separation $r$ is given by:

  \begin{equation}
	V_{\text{RK}} (r) = \begin{cases}
	  \frac{k e^2}{\kappa \rho_0} \left[ \ln \left( \frac{r}{2 \rho_0} \right) + \gamma \right]	&	r \ll \rho_0 \vspace{0.1cm} \\
	  - \frac{k e^2}{\kappa r}	&	r \gg \rho_0
	\end{cases}.
	\label{eq:rkasymptotic}
  \end{equation}
  where $\gamma$ is Euler's constant.

  % The derivation of the RK potential some four decades ago was driven by the need for an accurate model of the Coulomb interaction between charged particles in semiconductor quantum wells, which are layered semiconductor structures with thicknesses on the order of $\approx 50-100$ nm.
  % Therefore, $\rho_{0}$ was originally expressed in terms of the thickness $l$ of the semiconductor film and the static dielectric constant $\epsilon$ of the bulk material.
  % With the rise of 2D materials, the RK potential, and the screening length in particular, was re-examined.
  % For atomically thin materials, where the thickness of a monolayer is roughly 3-7 \AA, an alternative expression for the RK screening length $\rho_0$ was derived~\cite{Cudazzo2011,Berkelbach2013} in the limit of infinitesimal layer thickness,

  % Eq.~\eqref{eq:rho0chi2d} is often used to calculate $\rho_0$ in 2D materials, and we will follow this convention in our own calculations \textendash{} the value of $\chi_{2D}$ will be presented alongside the other input parameters in Sec.~\ref{sec:approach}.

  It was determined theoretically~\cite{Asahina1984} that the static dielectric constant of phosphorene is anisotropic, $\epsilon^x = 12.5,~\epsilon^y = 10.2$, and a more recent \textit{ab-initio} study~\cite{Rodin2014} likewise found that the 2D polarizability $\chi_{2D}$ was anisotropic, $\chi_{2D}^{x}=0.42$ nm, $\chi_{2D}^y = 0.397$ nm.
  However, the authors of Ref.~\onlinecite{Rodin2014} found that if $\chi_{2D}^x \approx \chi_{2D}^{y}$, one can approximate the 2D polarizability as isotropic by taking the average $\chi_{2D} = (\chi_{2D}^x + \chi_{2D}^y)/2$ without changing the results significantly, and we will employ the same approach here.

  While the RK potential has been applied to the study of indirect excitons in 2D heterostructures~\cite{Fogler2014,Berman2017,Berman2017a,Brunetti2018a,Brunetti2018b}, it is still common to use the Coulomb potential to study these systems.
  Therefore, for indirect excitons, we will solve the 2D Schr\"{o}dinger equation using both the RK and Coulomb potentials and compare the results.
  The Coulomb potential describing the interaction between spatially separated electrons and holes in a PHP HS can be written as:

  \begin{equation}
	V_{\text{C}} \left( \mathbf{r} \right) \equiv V_{\text{C}} \left( r \right) = -\frac{k e^2}{\epsilon' \sqrt{x^2 + y^2 + D^2}},
	\label{eq:vcoul}
  \end{equation}
  where the dielectric constant $\epsilon'$ takes the value of the environment, i.e. $\epsilon' = \kappa$.

  Now, the eigenenergies $E$ and corresponding eigenfunctions $\psi(\mathbf{r})$ of the exciton are obtained by solving the Schr\"{o}dinger equation~\eqref{eq:relschro} with the RK potential~\eqref{eq:vkeld} and $D=0$ for direct excitons, or for indirect excitons with either the RK~\eqref{eq:vkeld} or Coulomb~\eqref{eq:vcoul} potentials and $D = l_{phos} + N_{\text{BN}} l_{\text{BN}}$.

\section{\label{sec:optics}Exciton Optical Absorption}

  Calculations of the optical properties of excitons in phosphorene employ well-established methods~\cite{Snokebook} for modeling the response of atomic-like systems to an incident electromagnetic (EM) wave of frequency $\omega$ and polarization $\hat{e}$.
This approach was successfully used to study optical transitions in excitons in semiconductor quantum wells~\cite{Lozovik1997}, and has recently been applied to excitons in 2D materials~\cite{Brunetti2018a,Brunetti2018b}.
  The optical transition energy $E_{tr}$ corresponds to the difference in energy between the initial state, $\psi_i$, and the final state, $\psi_f$, and must coincide with the energy of the incident EM wave, i.e.\ $E_{tr} = E_f - E_i = \hbar \omega$.

  % Although optical transitions are modeled as a two-state system with specific initial and final states, it is usually the case that any given initial state can transition to a multitude of possible final states, with transitions to different final states being characterized in general by different transition energies.
  The oscillator strength, $f_0$, is a dimensionless quantity which gives the relative strength of a particular optical transition.
  For the isotropic 2D exciton, $f_0$ is proportional to the exciton reduced mass $\mu$ and does not depend on the in-plane orientation of the linearly polarized EM wave~\cite{Snokebook}.
  Modifying the standard expression for $f_0$ to account for anisotropy, we consider and calculate two distinct oscillator strengths, $f_{0}^{j}$, which correspond to the oscillator strengths of optical transitions induced by linearly polarized light oriented along the $x$ and $y$ axes, respectively.
  The polarization-dependent oscillator strength is thus given by:

  \begin{equation}
	f_0^j = \frac{2 \mu^j (E_f - E_i) \lvert \langle \psi_f \vert \hat{e} \vert \psi_i \rangle \rvert^2}{\hbar^2}.
	\label{eq:f0xy}
  \end{equation}

  In Eq.~\eqref{eq:f0xy}, $\vert \psi_i \rangle$ and $\vert \psi_f \rangle$ are the wavefunctions of the initial and final states, respectively.
  The allowed and forbidden optical transitions for a particular polarization $\hat{e}$ can be determined by calculating $f_{0}^{j}$, or more specifically by computing the dipole transition matrix element, $\lvert \langle \psi_f \vert \hat{e} \vert \psi_i \rangle \rvert^2$, which represents the overlap integral between the initial and final wavefunctions when the initial state interacts with an external electric dipole moment.
  The dipole transition matrix element is zero for forbidden transitions and non-zero if the transition is allowed.
  For allowed transitions, the oscillator strength is positive under photon absorption ($E_f - E_i > 0$) and negative under photon emission ($E_f - E_i < 0$).

  A theoretical study~\cite{Rodin2014} of the eigenstates of excitons in phosphorene with the RK potential using both Gaussian and sinudoidal basis functions provides crucial insight into the allowed optical transitions of the anisotropic exciton.
  It is well known that for 2D-hydrogen-like systems with either the Coulomb (isotropic dielectric environment) or RK (thin semiconducting film in an inhomogeneous dielectric environment) potentials, the allowed optical transitions of linearly polarized light are strictly limited to those in which the angular momentum quantum number differs by 1 between the initial and final eigenstates~\cite{LandauLif,Rodin2014}.
  For the anisotropic exciton, on the other hand, the authors of Ref.~\onlinecite{Rodin2014} found that linearly polarized light can induce a transition between any two states in which the symmetry of the eigenfunction along the polarization axis changes from even to odd, or vice versa.
  For example, the ground state eigenfunction is even along both the $x$ and $y$ axes, $n_x = n_y = 0$.
  Therefore, linearly polarized light along the $y$ direction can induce a transition to any state which is odd with respect to $y$, that is, $n_y' = n_y + 1,3,5,\dots;~n_x' = n_x$, while $x$-polarized light can likewise induce a transition from the ground state to the eigenstates characterized by $n_x' = n_x + 1,3,5,\dots;~n_y' = n_y$.

  While the oscillator strength gives us insight into the relationship between the eigenstates of the system and their response to an external EM force, there is a related quantity, the absorption coefficient $\alpha(\omega)$, which describes how strongly a particular material absorbs light of a given frequency due to an optical transition specified by $f_{0}^{j}$.
  Let us consider the attenuation of an EM wave propagating through a homogeneous material.
  The intensity $I$ of an EM wave of frequency $\omega$ is a function of the propagation distance $z$, given by:

  \begin{equation}
	I(z,\omega) = I_0 e^{- \alpha(\omega) z},
	\label{eq:iwz}
  \end{equation}
  where $I_0$ is the initial intensity of the wave.
  Eq.~\eqref{eq:iwz} illustrates the physical meaning of the absorption coefficient $\alpha(\omega)$, i.e.\ it is the reciprocal of the propagation distance $z$ over which the intensity of the EM wave of frequency $\omega$ decreases by a factor $e$.
  The absorption coefficient of optical transitions in isotropic atomic systems is given by~\cite{Snokebook}

  \begin{equation}
	\alpha(\omega) = \left( \frac{\omega}{\omega_0 c} \frac{\pi e^2}{2 \epsilon_0 \sqrt{\kappa} \mu} \frac{n_X}{l_{eff}} f_0 \right) \left( \frac{(\Gamma/2)}{(\omega_0^2 - \omega^2)^2 + (\Gamma/2)^2} \right),
	\label{eq:alpha}
  \end{equation}
  where $\omega_0 = (E_f - E_i)/\hbar$ is the Bohr angular frequency of the optical transition, $c$ is the speed of light, $n_X$ is the 2D concentration of excitons in the system, $l_{eff}$ is the effective vertical spatial extent of the exciton wavefunction, and $\Gamma$ is the full width at half maximum of the optical transition, often referred to as the broadening, line broadening, or damping.

  The fraction $n_X/l_{eff}$ physically represents the 3D exciton denisty, which for direct excitons in ML phosphorene can be straightforwardly written as the 2D concentration divided by the ML thickness, $l_{eff} = l_{phos}$.
  Considering indirect excitons in a PHP HS with the same 2D concentration $n_X$, it follows that the 3D exciton density must be reduced compared to direct excitons, i.e., $l_{eff} > l_{phos}$.
  In this case we consider the electron and hole to be bound to their respective phosphorene monolayers, and that the wavefunctions of the electron and hole do not penetrate far into the surrounding $h$-BN, so that the indirect excitons are effective contained within the two phosphorene monolayers.
  As the EM wave passes through the PHP HS, it only interacts with the exciton in regions where the exciton wavefunction is appreciably non-zero, i.e.\ the interaction only occurs within the phosphorene monolayers themselves.
  Therefore, we use $l_{eff} = 2 l_{phos}$ for indirect excitons.

  Summing Eq.~\eqref{eq:alpha} over all possible optically induced transitions in a given material (that is, not restricted to excitonic transitions) yields the absorption spectrum, a thorough description of how strongly the material absorbs light of frequency $\omega$.
  Since we only consider a very limited subset of all possible optical transitions in phosphorene, let us focus on the scenario where the energy of the incident excitation is equal to the energy of the transition given by $f_{0}^{j}$, $\hbar \omega = E_f - E_i$, which corresponds to a local maximum in the absorption spectrum $\alpha(\omega)$:

  \begin{equation}
	\alpha^j(\omega=\omega_0) \equiv \alpha^j = \left( \frac{\pi e^2}{2 c \epsilon_0 \sqrt{\kappa} \mu^j} \frac{n_X}{l_{eff}} f_0^j \right) \left( \frac{2}{\Gamma} \right).
	\label{eq:alphasimplify}
  \end{equation}
  Eq.~\eqref{eq:alphasimplify} can be used to characterize how strongly a particular optical transition absorbs the incident excitation.
  Additionally, the value of $\alpha^{j}$ can be used to conveniently compare the relative strengths of different excitonic transitions.

  As previously stated, the anisotropic absorption coefficient, $\alpha^{j}$, describes the attenuation of an EM wave of frequency $\omega$ and polarization $\hat{e}$ as a propagates an arbitrary distance $z$ through a dielectric.
  2D materials, however, do not have arbitrary thickness \textendash{} indeed, 2D materials are noteworthy precisely because each monolayer has a well-defined thickness.
  Therefore, it would be instructive to consider how the intensity of the incoming EM wave is reduced due to the wave propagating a distance which corresponds exactly to the thickness of the phosphorene ML(s) occupied by the excitons.
  Recalling Eq.~\eqref{eq:iwz}, we now define the polarization-dependent absorption factor as $\mathcal{A}^{j} \equiv 1 - \left( I(z = l_{eff}, \omega = \omega_0)/I_0 \right)$, or:

  \begin{equation}
	\mathcal{A}^{j} = 1 - \text{exp} \left[ - \alpha^j l_{eff} \right] = 1 - \text{exp} \left[ - \left( \frac{\pi e^2 n_X}{2 c \epsilon_0 \sqrt{\kappa} \mu^j} f_0^j \right) \left( \frac{2}{\Gamma} \right) \right].
	\label{eq:afac}
  \end{equation}
  Eq.~\eqref{eq:afac} therefore gives the fractional decrease in the intensity of the EM wave as it propagates through one exciton layer (that is, one ML for direct excitons, or one PHP HS for indirect excitons), i.e.\ $\mathcal{A} = 0.01$ means that each exciton layer absorbs 1\% of the incident EM wave.
  The absorption factor is particularly convenient when comparing absorption between direct and indirect excitons, or between excitons in 2D materials with different thicknesses.

  Simplified forms of Eqs.~\eqref{eq:alphasimplify} and~\eqref{eq:afac} are given by Eqs.~\eqref{eq:ctildecalphacases}-~\eqref{eq:afacapprox} in Appendix~\ref{app:approach-scales}.

\section{\label{sec:approach}Computational Approach}

\subsection{\label{ssec:approach-params}Discussion of input parameters used in numerical calculations}

  Calculating the optical properties of the exciton using Eqs.~\eqref{eq:f0xy},~\eqref{eq:alphasimplify}, and~\eqref{eq:afac} requires the excitonic eigenenergies $E_n$ and eigenfunctions $\vert \psi_n \rangle$, which are obtained by solving the Schr\"{o}dinger equation~\eqref{eq:relschro} with either the RK potential~\eqref{eq:vkeld} (for both direct and indirect excitons) or the Coulomb potential~\eqref{eq:vcoul} (for indirect excitons only).
  The Schr\"{o}dinger equation takes as input parameters the anisotropic exciton reduced masses $\mu^x$ and $\mu^y$ and either the uniform dielectric constant $\epsilon'$ for the indirect exciton with the Coulomb potential, or the average environmental dielectric constant $\kappa$ and 2D polarizability $\chi_{2D}$ for either direct or indirect excitons with the RK potential.

  Numerical solution of the Schr\"{o}dinger equation using the aforementioned interaction potentials and input parameters is performed using the finite element method (FEM), which yields $N$ pairs of eigenenergies and eigenfunctions which are solutions to the Schr\"{o}dinger equation, corresponding to the $N$ most-strongly-bound states.
  The eigenenergies $E_n$ and eigenfunctions $\psi_n$, along with the appropriate anisotropic reduced mass $\mu^j$, are then used as input parameters to calculate the oscillator strength, $f_{0}^{j}$ according to Eq.~\eqref{eq:f0xy}.
  The corresponding polarization-dependent absorption coefficients $\alpha^{j}$ and absorption factors $\mathcal{A}^{j}$ can then be calculated using as inputs the oscillator strength, $f_{0}^{j}$, the anisotropic exciton reduced mass $\mu^j$, the 2D exciton concentration $n_X$, the phosphorene ML thickness $l$, environmental dielectric constant $\kappa$ (or $\epsilon'$ for the Coulomb potential), and the excitonic optical broadening $\Gamma$.

\begin{table}[ht]
  \centering
  \caption{%
	All input parameters used in numerical calculations.
	Each set of masses, denoted $\mu_i$, $i=a,b,c,d$, was taken from the corresponding reference.
	The anisotropic exciton reduced masses $\mu^{j}$, calculated using the corresponding effective charge carrier masses $m_{i}^{j}$, are printed in bold to aid the eye.
	The next three columns denote the following quantities: the phosphorene ML thickness, $l_{phos}$; the 2D polarizability, $\chi_{2D}$; and the 2D exciton concentration, $n_X$.
	The column titled ``Env.'' denotes the four dielectric environments for which we calculate the properties of direct excitons in ML phosphorene: freestanding, i.e.\ suspended in vacuum (FS); supported on either an SiO$_2$ substrate (SS) or $h$-BN substrate (HS) with the top of the ML exposed to air or vacuum (``uncapped''); and encapsulated by $h$-BN on the top and bottom (HE).
	Each of the four environments are associated with a particular value for $\kappa$ and $\Gamma$, given in the following two columns, respectively.
	The value of $\Gamma$ for FS, SS, and HS was chosen based on Refs.~\onlinecite{Yang2015a,Liu2014a,Wang2015b}, while the value for HE is based on Refs.~\onlinecite{Cadiz2017,Robert2017,Horng2018}.
	Additional discussion of these quantities are given in the text below.
  \label{tab:mutab}
}
  \begin{tabular}{ccccccccccccc}
	\toprule[0.05em]\toprule[0.05em]
	{}	& $m_e^x$	& $m_h^x$ 	& $\mu^x$	& $m_e^y$	& $m_h^y$	& $\mu^y$	& $l_{phos}$ [nm]~\cite{Kumar2016}	& $\chi_{2D}$ [nm]~\cite{Rodin2014}	& $n_X$ [m$^{-2}$]~\cite{You2015}	& Env. & $\kappa$ 	& $\Gamma$ [s$^{-1}$]	\\ \midrule[0.05em]
	$\mu_{a}$~\cite{Peng2014}	& 0.16	& 0.15		& \textbf{0.0630}	& 1.24		& 4.92		& \textbf{0.968}	& \multirow{4}{*}{0.541} & \multirow{4}{*}{0.41} & \multirow{4}{*}{$5 \times 10^{15}$}	& FS 	& 1 		& $10^{14}$ \\
	$\mu_{b}$~\cite{Tran2014a}	& 0.1	& 0.2		& \textbf{0.0667}	& 1.3		& 2.8		& \textbf{0.888}	& {} & {} 			&	{}									& SS 	& 2.4		& $10^{14}$ \\
	$\mu_{c}$~\cite{Paez2016}	& 0.199	& 0.1678	& \textbf{0.0910}	& 0.7527	& 5.35		& \textbf{0.660}	& {} & {}			&	{}									& HS 	& 2.945		& $10^{14}$ \\
	$\mu_{d}$~\cite{Qiao2014}	& 0.17	& 0.15		& \textbf{0.0797}	& 1.12		& 6.35		& \textbf{0.952}	& {} & {}			&	{}									& HE 	& 4.89 		& $10^{13}$	\\ \bottomrule[0.05em]
  \end{tabular}
\end{table}

  The standard values of these input parameters are given in Table~\ref{tab:mutab} \textendash{} unless otherwise noted, all subsequent results were obtained using these values.
  Whereas the sets of carrier masses $\mu_i$, $i=a,b,c,d$, were straightforwardly taken from the corresponding references, some additional discussion of the other input parameters is necessary.

  The phosphorene ML thickness, $l_{phos}$, obtained via \textit{ab-initio} calculations in Ref.~\onlinecite{Kumar2016}, agrees well with theoretical results from other works, namely $0.53$ nm~\cite{Wang2015b} and $0.6$ nm~\cite{Liu2014a} \textendash{} we note that these last two references also measured the ML thickness using atomic force microscopy (AFM), obtaining values of $0.85$ and $0.7$ nm, respectively, but the authors themselves note that AFM measurements tend to over-estimate ML thickness.
  The 2D polarizability, $\chi_{2D}$, was calculated from first-principles in Ref.~\onlinecite{Rodin2014} and agrees well with the value of $0.38$ nm, also obtained from first-principles in Ref.~\onlinecite{Prada2015}.

  The 2D exciton concentration, $n_X$, differs from the previous quantities in that it is not a material property that can be definitively measured or calculated \textendash{} instead, $n_X$ depends mainly on the excitation intensity, that is, a high-intensity laser will excite a higher concentration of excitons than a low-intensity laser.
  Therefore, it is reasonable to expect that $n_X$ can and will vary significantly between experimental configurations, and even from one trial to the next.
  Instead of exhaustively considering a wide range of possible values of $n_X$, we instead choose one value of $n_X$ which is representative of a typical experiment to use throughout our calculations.
  One recent study~\cite{Surrente2016} of exciton-exciton annihilation rates in a phosphorene ML found that exciton-exciton annhiliation becomes the dominant recombination mechanism (as opposed to e.g.\ thermal decomposition or radiative recombination) at an exciton concentration of about $6.1 \times 10^{16}$ m$^{-2}$.
  While this value is not representative of a typical experiment, it may still be helpful to consider as an upper bound.
  Lacking an appropriate result from experimental studies in phosphorene, we turn instead to excitons in TMDCs, where we find a reasonable value of $n_X$ in Ref.~\onlinecite{You2015}, which studied excitons in a WSe$_{\text{2}}$ ML.

  Based on previous optical studies of excitons in 2D materials, we will use two values of $\Gamma$.
  It is important to note that the $\Gamma$ obtained from experimental measurements is, like $n_X$, dependent on several external factors, including but not limited to: the sample temperature; the presence of structural defects within the sample; and/or surface contaminants at either the substrate/monolayer interface or the monolayer/air interface.
  These confounding variables can significantly alter the observed optical properties of the material, especially the presence of defects and contaminants which may be difficult to identify, characterize, isolate, and prevent.

  In choosing a value for $\Gamma$, we therefore adopt a similar approach to our choice of a value for $n_X$ \textendash{} we will choose a value for $\Gamma$ which is generally appropriate for the system in question, but need not correspond exactly to one particular observed value.
  Many experimental PL/absorption studies of excitons in phosphorene are conducted with the phosphorene ML placed on a substrate (typically SiO$_2$), while the opposite side of the ML is left exposed to the atmosphere.
  These studies all observed significant broadening of the excitonic emission/absorption peak, with reported values of 70 meV~\cite{Yang2015a}, 100 meV~\cite{Liu2014a}, and 150 meV~\cite{Wang2015b}.
  Therefore, when calculating $\alpha$ and $\mathcal{A}$ for FS or uncapped phosphorene (that is, on an SiO$_2$ or $h$-BN substrate), we will use the value $\Gamma = (70~\text{meV})/\hbar \approx 10^{14}$ s$^{-1}$.
  For phosphorene encapsulated by $h$-BN, we again turn to similar studies on the TMDCs, where large excitonic broadening was observed in uncapped TMDC samples at room temperature~\cite{Steinhoff2014,Molina-Sanchez2013,Mak2010}, but encapsulating the TMDC with $h$-BN was found to drastically reduce the excitonic linewidths to their cryogenic limit~\cite{Cadiz2017,Robert2017,Horng2018}, $\Gamma = (11~\text{meV})/\hbar \approx 10^{13}$ s$^{-1}$.
  Results for $\alpha$ and $\mathcal{A}$ for direct excitons will therefore be presented using two different values of $\Gamma$, depending on the dielectric environment, while for indirect excitons only $\Gamma = 10^{13}$ s$^{-1}$ will be used since only $h$-BN encapsulation of the PHP HS is considered.

\section{\label{sec:results}Results of Calculations}

  \subsection{\label{ssec:directresults}Direct Excitons}

  In this Section we present the results of our calculations of the eigenenergies and optical properties of the direct exciton using the input parameters listed in Table~\ref{tab:mutab}, focusing in particular on how our results change depending on the four sets of masses $\mu_{i}$ and the four dielectric environments denoted by FS, SS, HS, and HE\@.
  The notation $X[i,k]$ will be used as a shorthand for ``the value of the quantity $X$ calculated using the set $\mu_{i}$ in the dielectric environment $k \in \text{[FS, SS, HS, HE]}$'', i.e.\ $f_{0}^{y}[a,\text{FS}]$ means ``the value of $f_{0}^{y}$ in FS phosphorene calculated using $\mu_{a}$.''
  For the input parameters that produce the minimum or maximum value of a particular quantity, a $min$ or $max$ subscript will be added to the corresponding parameter, i.e.\ $E_{b}[d_{max},k]$ means that $\mu_d$ yields the maximum value of $E_b$ for the given $k$.
  The percent difference between the maximum and minimum values, with respect to e.g.\ the $\mu_i$, of a particular quantity will be denoted with a $\%$ subscript on the parameter, i.e.\ $E_{b}[i_{\%},k]$.
  Averaging a quantity over a set of parameters will be denoted by the subscript \textit{avg.}, as in $E_{b}[i_{avg.},k]$.
  If $i$ or $k$ have been previously established in context, either index may be omitted from the notation $X[i,k]$.

  % \begin{figure}[ht]
  %   \centering
  %   \includegraphics[width=0.7\columnwidth]{phos-comb-0717-dir-En-allenv-ebars.pdf}
  %   \caption{%
  %     Eigenenergies of the direct exciton in four different dielectric environments: freestanding (FS) (inset), supported on an SiO$_2$ (SS) or $h$-BN (HS) substrate, and encapsulated by $h$-BN (HE).
  %     For each data set, the values corresponding to the four $\mu_i$ are shown by the data point (mean value) and the interval markers, whose boundaries correspond to the maximum and minimum values  of the four $\mu_i$.
  %   }
  %   \label{fig:dir-En-allmus}
  % \end{figure}

  \begin{table}
	\centering
	\caption{%
	  Eigenenergies of the direct exciton in four dielectric environments: freestanding (FS), supported on an SiO$_2$ (SS) or $h$-BN (HS) substrate, and encapsulated by $h$-BN (HE).
	}
	\begin{tabular}{lrlrlrlrl}
	  \midrule[0.05em]\midrule[0.05em]
	  {}				&	\multicolumn{2}{c}{$E_n[\text{FS}]$, [meV]}	&	\multicolumn{2}{c}{$E_n[\text{SS}]$, [meV]}	&	\multicolumn{2}{c}{$E_n[\text{HS}]$, [meV]}	&	\multicolumn{2}{c}{$E_n[\text{HE}]$, [meV]}	\\ \midrule[0.05em]
	  $\vert n \rangle$	&	Min		&	Max			&	Min			&	Max		&	Min			&	Max		&	Min		&	Max			\\ \midrule[0.05em]
	  $\vert 1 \rangle$ &	718.7	&	753.3		&	381.7		&	407.5	&	317.8		&	341.1	&	187.9	&	204.7		\\ 
	  $\vert 2 \rangle$ &	478.6	&	508.1		&	202.2		&	221.9	&	156.3		&	173.2	&	74.32	&	84.61		\\ 
	  $\vert 3 \rangle$ &	377.6	&	408.7		&	144.7		&	162.2	&	109.5		&	123.7	&	50.31	&	57.78		\\ 
	  $\vert 4 \rangle$ &	310.9	&	330.3		&	105.6		&	117.7	&	76.97		&	87.22	&	31.97	&	37.94		\\ 
	  $\vert 5 \rangle$ &	272.4	&	300.0		&	88.40		&	101.3	&	63.60		&	74.25	&	26.96	&	31.62		\\ 
	  $\vert 6 \rangle$ &	250.4	&	281.3		&	80.36		&	94.27	&	58.38		&	69.04	&	24.70	&	29.61		\\ \midrule[0.05em]
	  \bottomrule
	\end{tabular}
	\label{tab:dir-En-minmax}
  \end{table}

  In Table~\ref{tab:dir-En-minmax}, the eigenenergies of the direct exciton in a phosphorene ML are calculated for four different dielectric environments and for each of the four sets of $\mu_i$ from Table~\ref{tab:mutab}.
  The binding energies follow the relation $E_{b}[d,k] > E_{b}[c,k] > E_{b}[b,k] > E_{b}[a,k]$ for all dielectric environments $k$.
  The choice of the $\mu_i$ does not change $E_b$ significantly, though the difference between the minimum and maximum $E_b$ increases with $\kappa$, e.g.\ we find $E_{b}[i_{\%},\text{FS}] \approx 5\%$, while $E_{b}[i_{\%},\text{HE}] \approx 8.5\%$.
  This percent difference also increases for higher excited states.
  Of course, the parameter which most significantly changes $E_b$ is $\kappa$, where $E_{b}[i,\text{FS}] \approx 2 E_{b}[i,\text{(SS,HS)}]$ and $E_{b}[i,\text{SS}] \approx 2 E_{b}[i,\text{HE}]$.

  In addition, our results for the binding energy of the direct exciton shown in Table~\ref{tab:dir-En-minmax} agree very well with previously reported results in a variety of different dielectric environments.
  The binding energy of the direct exciton in a FS phosphorene ML has been calculated via \textit{ab-initio} methods on several occasions \textendash{} prior calculations vary between 700-850 meV~\cite{Cakir2014b,Caklr2015,Tran2014,Choi2015,Rodin2014,Seixas2015}, which agrees quite well with our average value of about 740 meV, considering the previous results were obtained using a variety of methods and, therefore, a variety of input parameters.
  Regarding phosphorene on an SiO$_2$ substrate, the direct exciton binding energy was theoretically calculated to be about 400 meV~\cite{Rodin2014}, which is within the range shown in Table~\ref{tab:dir-En-minmax}.
  Another experimental study of phosphorene on an SiO$_2$ substrate~\cite{Yang2015a} determined the binding energy to be about 300 meV, while a separate experimental investigation performed around the same time~\cite{Wang2015b} obtained a surprisingly high value of 900 meV.
  Finally, in Ref.~\onlinecite{Rodin2014}, where the electron-hole interaction was also modeled using the RK potential, the direct exciton binding energy was calculated to be about 200 meV for $\kappa = 5$, which falls within our calculated range for HE ($\kappa = 4.89$).

  \begin{figure}[ht]
	\centering
	\includegraphics[width=0.65\columnwidth]{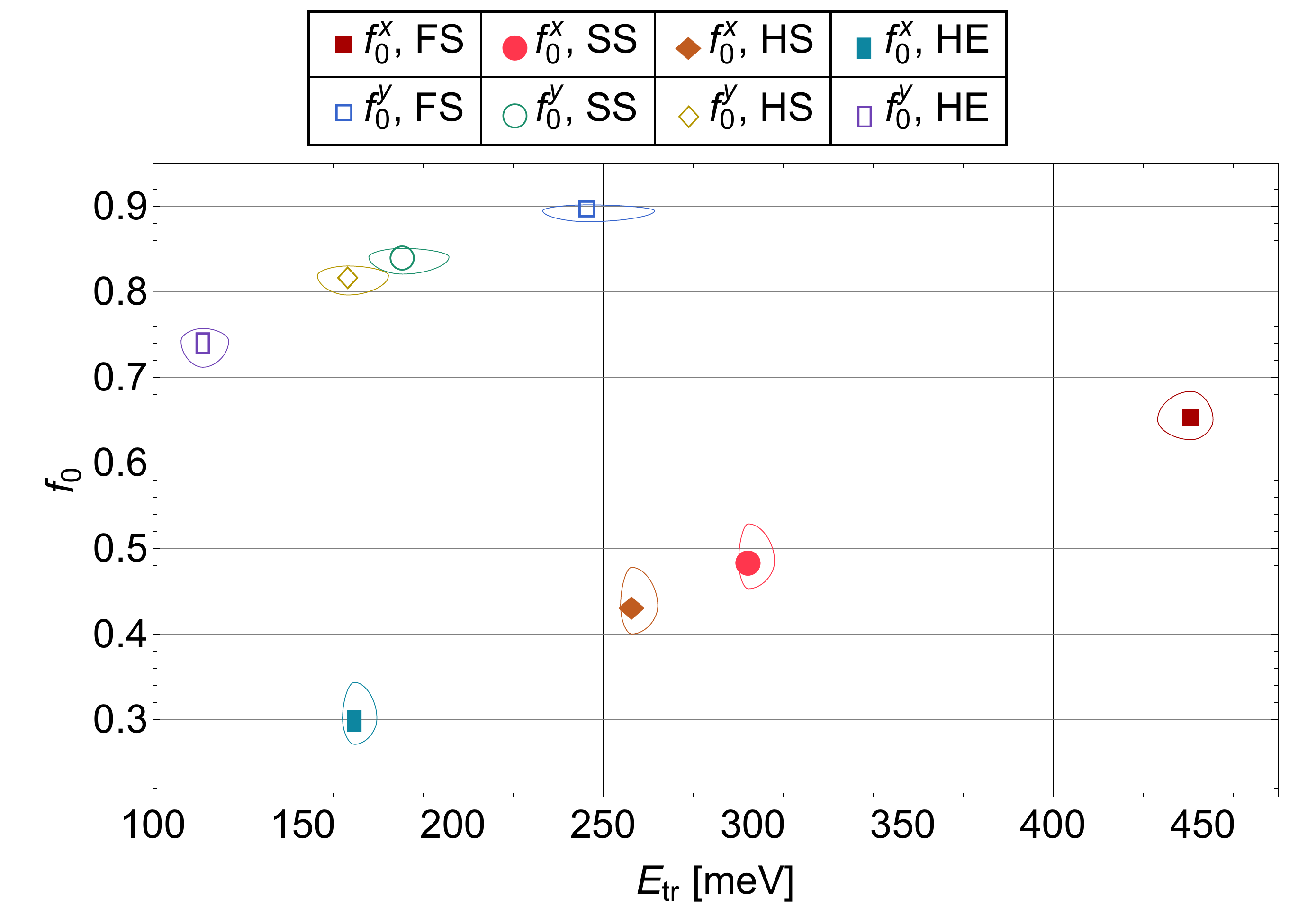}
	\caption{%
	  Relationship between $f_{0}^{j}$ and $E_{tr}^{j}$ for the first allowed optical transition (i.e.\ the $x$ or $y$ transition with the smallest transition energy) under $x$- and $y$-polarized excitations, shown by solid and open markers, respectively.
	  The plot marker denotes the values of $E_{tr}^{j}$ and $f_{0}^{j}$ averaged over the four $\mu_{i}$, while the axes of the ellipses around each data point correspond to the minimum and maximum values of $E_{tr}^{j}$ and $f_{0}^{j}$.
	}
	\label{fig:dir-foxy-vs-etr-allmus-allenvs}
  \end{figure}

  The oscillator strengths of the first allowed optical transitions for $x$- and $y$-polarized light are shown for all four dielectric environments in Fig.~\ref{fig:dir-foxy-vs-etr-allmus-allenvs}.
  In particular, $f_{0}^{y}$ (shown with the open markers) refers to the $\left( 0,0 \right) \to (0,1)$ transition and $f_{0}^{x}$ (shown by the solid markers) refers to the $(0,0)\to(1,0)$ transition.
  Since both $E_{tr}^{j}$ and $f_{0}^{j}$ depend on $\mu_i$, both quantities are averaged across the four $\mu_i$ and the average value is denoted by the plot marker.
  The major and minor axes of the ellipses encircling each data point mark the minimum and maximum values of $E_{tr}^{j}$ and $f_{0}^{j}$.

  From Fig.~\ref{fig:dir-foxy-vs-etr-allmus-allenvs}, we see that both $f_{0}^{j}$ and $E_{tr}^{j}$ are decreasing functions of $\kappa$.
  The effect of anisotropy is also evident in the relative magnitudes of $f_{0}^{x}$ and $f_{0}^{y}$, where $f_{0}^{y}[i,k] \gg f_{0}^{x}[i,k]$ for all $k$, and furthermore, $f_{0}^{y}[i,\text{HE}] > f_{0}^{x}[i,\text{FS}]$

  \begin{table}[ht]
	\centering
	\caption{%
	  Calculated ratios $\widetilde{f}_{0}^{j}[i,k] \equiv f_{0}^{j}/\mu_{i}^{j}$, averaged over the four $\mu_i$, as well as the absorption coefficient scale factor, $\widetilde{C}_{D}$, from Eq.~\eqref{eq:ctildecalphacases}.
	  For tabulated values of the $\widetilde{f}_{0}^{j}$ for each $\mu_i$, see Appendix~\ref{app:approach-scales}, Table~\ref{tab:dirf0muratiosfull}.
	  The units of $\widetilde{f}_{0}^{j}$ are [m$_{0}^{-1}$].
	}
	\begin{tabular}{c c c c c c c c c c}
	  \toprule[0.05em]\toprule[0.05em]
	  Env.	& $\widetilde{f}_{0}^{x}[i_{avg.}]$	& $\widetilde{f}_{0}^{y}[i_{avg.}]$	& $\widetilde{C}_{D}$ [$\times 10^{6}$ m$^{-1}$] \\ \midrule[0.05em]
	  FS	& 8.812							& 1.056							& $3.082$\\
	  SS	& 6.543							& 0.991							& $1.990$\\
	  HS	& 5.834							& 0.965							& $1.796$\\
	  HE	& 4.05							& 0.874							& $13.93$\\ \bottomrule[0.05em]
	\end{tabular}
	\label{tab:dirf0muratios}
  \end{table}

  Table~\ref{tab:dirf0muratios} shows the ratios $\widetilde{f}_{0}^{j}[i_{avg.},k]$ along with the corresponding $\widetilde{C}_{D}[k]$.
  Following the procedure outlined in Appendix~\ref{app:approach-scales}, the average absorption coefficient with respect to the $\mu_i$ can be easily calculated as $\alpha^{j}[i_{avg.},k] = \widetilde{C}_{D}[i_{avg.},k] \widetilde{f}_{0}^{j}[i_{avg.},k]$.

  Interestingly, whereas $f_{0}^{y}[i,k] > f_{0}^{x}[i,k]$ for all $i$ and $k$, we find that the opposite is true for $\alpha^{x}$ and $\alpha^{y}$.
  The reason for this can be seen from Tables~\ref{tab:mutab} and~\ref{tab:dirf0muratios} \textendash{} although $f_{0}^{y}$ can exceed $f_{0}^{x}$ by anywhere between about 30\% (in FS) and over 100\% (in HE), $\mu_{i}^{x}$ can be more than an order of magnitude larger than $\mu_{i}^{y}$, so that the ratio $\widetilde{f}_{0}^{x}$ is always greater than $\widetilde{f}_{0}^{y}$, and hence $\alpha^{x}$ will always be larger than $\alpha^{y}$.

  A prior study~\cite{Xia2014a} of the optical absorption and PL properties of ML phosphorene found that exciton-forming excitations were much more strongly absorbed if the excitation was polarized along $x$ than along $y$.
  While the underlying theory of exciton-forming transitions differs substantially from the treatment of intra-excitonic transitions, it is plausible that in both cases, the fact that $\mu^{x}$ is much smaller than $\mu^{y}$ leads to enhanced absorption of $x$-polarized light.
  In other words, the amplitude of the oscillatory response of the exciton to an $x$-polarized driving force is much larger than the amplitude of oscillations induced by a $y$-polarized excitation, due to the fact that $\mu^{y} \gg \mu^{x}$.

  From Eqs.~\eqref{eq:alphasimplify} and~\eqref{eq:afac}, we see that $\alpha^{j}$ and $\mathcal{A}^{j}$ are inversely proportional to $\sqrt{\kappa}$.
  At the same time, we find that $\widetilde{f}_{0}^{y}[i,\text{FS}] \approx 1.25 \widetilde{f}_{0}^{y}[i,\text{HE}]$, while $\widetilde{f}_{0}^{x}[i, \text{FS}] > 2 \widetilde{f}_{0}^{x}[i, \text{HE}]$.
  Combining these two trends and assuming for the moment that $\Gamma$ remains the same in all dielectric environments, we would expect a significant change in both $\alpha^{x}$ and $\alpha^{y}$, approximately $\alpha^{x}[i,\text{FS}] \approx 5 \alpha^{x}[i,\text{HE}]$ and $\alpha^{y}[i,\text{FS}] \approx 3 \alpha^{y}[i,\text{HE}]$.
  Assuming instead that $h$-BN encapsulation significantly reduces $\Gamma$, we find that $\alpha^{x}[\text{HE}]$ is about twice as large as $\alpha^{x}[\text{FS}]$, while $\alpha^{y}[\text{HE}]$ is greater than $\alpha^{y}[\text{FS}]$ by nearly a factor of four.
  The aborption factor $\mathcal{A}^{j}$ reveals the significant difference in excitonic optical activity between the two polarization directions and the four dielectric environments, where we obtain $\mathcal{A}^{x}[i_{avg.},k]$ as $1.36\%,~0.68\%,~0.55\%,~3.0\%$ and $\mathcal{A}^{y}[i_{avg.},k]$ as $0.017\%,~0.011\%,~0.009\%,~0.066\%$, for $k = $ FS, SS, HS, HE, respectively.

  Comparing the optical quantities $E_{tr}^{j}$, $f_{0}^{j}$, $\alpha^{j}$, and $\mathcal{A}^{j}$ for all $\mu_{i}$ and across all dielectric environments, we observe some general trends.
  Let us first address the quantities related to the $y$-transitions, followed by the $x$-related quantities.

  The optical transition energies follow the relation $E_{tr}^{y}[c,k] > E_{tr}^{y}[d,k] > E_{tr}^{y}[b,k] > E_{tr}^{y}[a,k]$ in all $k$.
  Curiously, the oscillator strengths in FS phosphorene are reversed compared to the transition energies, i.e.\ $f_{0}^{y}[a,\text{FS}] > f_{0}^{y}[b,\text{FS}] > f_{0}^{y}[d,\text{FS}] > f_{0}^{y}[c,\text{FS}]$, though we instead obtain $f_{0}^{y}[a,k'] > f_{0}^{y}[d,k'] > f_{0}^{y}[b,k'] > f_{0}^{y}[c,k']$ for $k' = \text{(SS, HS, HE)}$.
  The ordering with respect to the $\mu_i$ is reversed again for the $\widetilde{f}_{0}^{y}$, and therefore for $\alpha^{y}$ and $\mathcal{A}^{y}$ as well, i.e.\ $\widetilde{f}_{0}^{j}[c,k] > \widetilde{f}_{0}^{j}[b,k] > \widetilde{f}_{0}^{j}[d,k] > \widetilde{f}_{0}^{j}[a,k]$, for all $k$.
  Recalling from Table~\ref{tab:mutab} that $\mu_{a}^{y} > \mu_{d}^{y} > \mu_{b}^{y} > \mu_{c}^{y}$, it appears that $f_{0}^{j}$ is an increasing function of $\mu^{j}$, while $E_{tr}^{j}$ is a decreasing function of $\mu^{j}$.
  % Now, in all environments, it is true that $\mu_{a}$ produces both the smallest $E_{tr}^{y}$ and the largest $f_{0}^{y}$, while $\mu_{c}$ yields the largest $E_{tr}^{y}$ and the smallest $f_{0}^{y}$, but the inconsistent ordering for $f_{0}^{y}[b]$ and $f_{0}^{y}[d]$ complicate the relationship between $\mu_{i}^{y}$, $E_{tr}^{y}$, and $f_{0}^{y}$.
  % Indeed, from Table~\ref{tab:dirf0muratiosfull} we see that $\widetilde{f}_{c}^{y} > \widetilde{f}_{b}^{y} > \widetilde{f}_{d}^{y} > \widetilde{f}_{a}^{y}$ which shares the same maximum and minimum as the $E_{tr}^{y}$, while being the reverse of the $\mu_{i}$ and $f_{0}^{y}[\text{SS, HS, HE}]$.

  In contrast to the $y$-polarized quantities, whose relative magnitudes were constant across the four dielectric environments for $E_{tr}^{y}$ and $\widetilde{f}_{0}^{y}$ but were inconsistent in $f_{0}^{y}$, we find that the ordering of the $E_{tr}^{x}[i,k]$ is different for each $k$, while the relative magnitudes of both $f_{0}^{x}$ and $\widetilde{f}_{0}^{x}$ are consistent for all $k$.
  The transition energies $E_{tr}^{x}$ show significant variation between different environments, with the only constant being that $E_{tr}^{x}[d]$ is always the largest value.
  Whereas $E_{tr}^{x}[d,\text{FS}] > E_{tr}^{x}[a,\text{FS}] > E_{tr}^{x}[b,\text{FS}] > E_{tr}^{x}[c,\text{FS}]$, we find that the relative magnitude of $E_{tr}^{x}[c]$ increases as the dielectric screening increases, while $E_{tr}^{x}[a]$ decreases relative to the other values, such that $E_{tr}^{x}[d,\text{HE}] > E_{tr}^{x}[c,\text{HE}] > E_{tr}^{x}[b,\text{HE}] > E_{tr}^{x}[a,\text{HE}]$.
  The oscillator strengths, on the other hand, follow the same order as the $\mu_{i}^{x}$ themselves, i.e.\ $\mu_{c}^{x} > \mu_{d}^{x} > \mu_{b}^{x} > \mu_{a}^{x}$, for all dielectric environments $k$.
  Additionally, for all $k$, the ordering of the $\widetilde{f}_{0}^{x}$, $\alpha^{x}$, and $\mathcal{A}^{x}$ is reversed with respect to the $\mu_{i}^{x}$.

  These observations suggest that while the optical properties corresponding to a particular excitation polarization $\hat{e}$ are primarily determined by the corresponding $\mu^{j}$, these quantities also exhibit some dependence on the opposite $\mu^{j' \neq j}$, stemming from the dependence of the optical properties on the excitonic ground state, whose properties must represent both $\mu^x$ and $\mu^y$.
  Considering the uniquely strong response of the material properties of phosphorene, e.g.\ the anisotropic effective charge carrier masses, to external stimuli such as mechanical strain~\cite{Cakir2014b,Wang2015b}, the preceeding analysis should prove useful in guiding future efforts to engineer phosphorene MLs with specific optical properties.

\subsection{\label{ssec:indirectresults}Indirect Excitons}

  In this Section we present and analyze the same calculated quantities as for the direct exciton, now for the indirect exciton in a PHP HS\@.
  All quantities were calculated by solving the Schr\"{o}dinger equation with both the RK and Coulomb potentials, for different interlayer separations corresponding to $N_{\text{BN}} = 1,2,3,\dots,8$.
  We will adapt the notation $X[i,k]$ used in the previous section to accomodate the different input parameters for the indirect exciton.
  Here, the notation $X[i,p,N_{\text{BN}}]$ will denote ``the quantity $X$ calculated using $\mu_{i}$, the potential $p = (\text{RK, C})$, and interlayer separation $N_{\text{BN}}$.''

  \begin{figure}[ht]
	\centering
	\includegraphics[width=0.75\columnwidth]{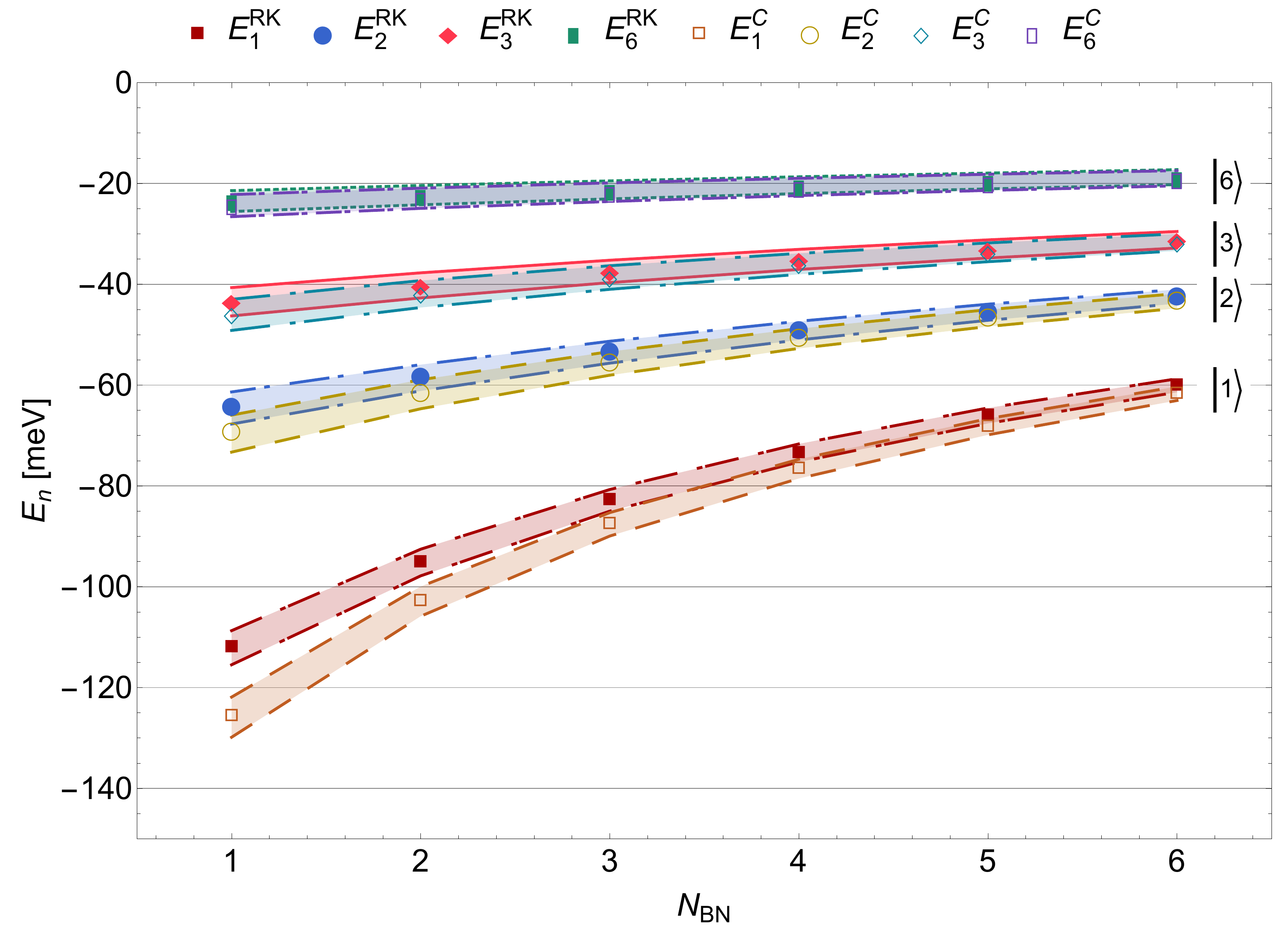}
	\caption{%
	  Comparison of the eigenstates $\vert 1 \rangle,\vert 2 \rangle ,\vert 3 \rangle$, and $\vert 6 \rangle $ for the indirect exciton, calculated using both the RK (solid markers) and Coulomb (open markers), as a function of the number of $h$-BN monolayers separating the phosphorene monolayers, $N_{BN}$.
	  The states $\vert 4 \rangle $ and $\vert 5 \rangle $ are not shown due to overlap with the $\vert 3 \rangle $ and $\vert 6 \rangle $ states.
	}
	\label{fig:ind-en-vs-nbn-rkc}
  \end{figure}

  In Fig.~\ref{fig:ind-en-vs-nbn-rkc}, we plot the dependence of the indirect exciton eigenenergies $E_{n}$, $n=1,2,3,6$, on the interlayer separation, $N_{\text{BN}}$, where all $E_n$ were calculated for both the RK (solid markers) and Coulomb (open markers) interaction potentials.
  Calculations were performed for all four $\mu_i$.
  The plot marker denotes the average value of the $\mu_i$, while the boundaries of the shaded regions denote the minimum and maximum values.

  The difference between the RK and Coulomb potentials is significant only for the first couple eigenstates at small interlayer separations.
  We find that the percent difference between the RK and Coulomb potentials decreases as $N_{\text{BN}}$ increases, that is, $E_b[i, p_{\%}, N_{BN}] \approx 11\%,~7.7\%,~5.5\%,\dots,~2\%$ for $N_{\text{BN}} = 1,~2,~3,\dots,~8$.
  The excited state energies follow a similar trend, where we find $E_{2}[i, p_{\%}, (1,8)] \approx (7.5\%,~1.5\%)$.
  As shown in Eq.~\eqref{eq:rkasymptotic} when the relative separation $\lvert \mathbf{r} \rvert$ exceeds the screening length, $\rho_0 = (2 \pi \chi_{2D})/\kappa$, the RK potential converges to the Coulomb potential.
  For a PHP HS with $\kappa = 4.89$ and $\chi_{2D} = 0.41$ nm, we calculate $\rho_0 = 0.526$ nm.
  Therefore, one would expect the RK and Coulomb potentials to converge as the total electron-hole separation exceeds 0.526 nm.
  Considering that $l_{\text{BN}} = 0.333$ nm, it is unsurprising that the indirect exciton binding energies for the RK and Coulomb potentials start to overlap as $N_{\text{BN}}>2$.
  The convergence of the excited state eigenenergies is also the result of increasing electron-hole separation, since the average separation of a two-particle bound state increases as progressively higher excited states are accessed.

  As with the direct exciton, the choice of $\mu_{i}$ does not significantly change the indirect exciton binding energy \textendash{} for example, we calculate $E_{b}[i_{\%}, p, 1] \approx 6\%$, decreasing to about $E_{b}[i_{\%}, p, 8] \approx 4\%$.
  Although the value of $\kappa$ is the same, the indirect exciton binding energy is reduced by about 40\% compared to the direct exciton in HE due to the increased electron-hole separation in the PHP HS, from $E_b[i,HE] \approx 200$ meV to $E_b[i, p, 1] \approx 120$ meV.

  \begin{figure}[ht]
	\centering
	\includegraphics[width=0.7\columnwidth]{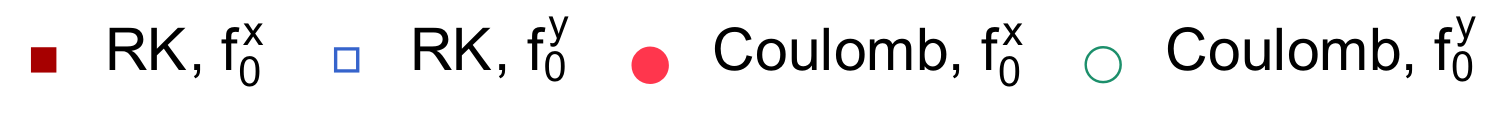}\\
	\includegraphics[width=0.75\columnwidth]{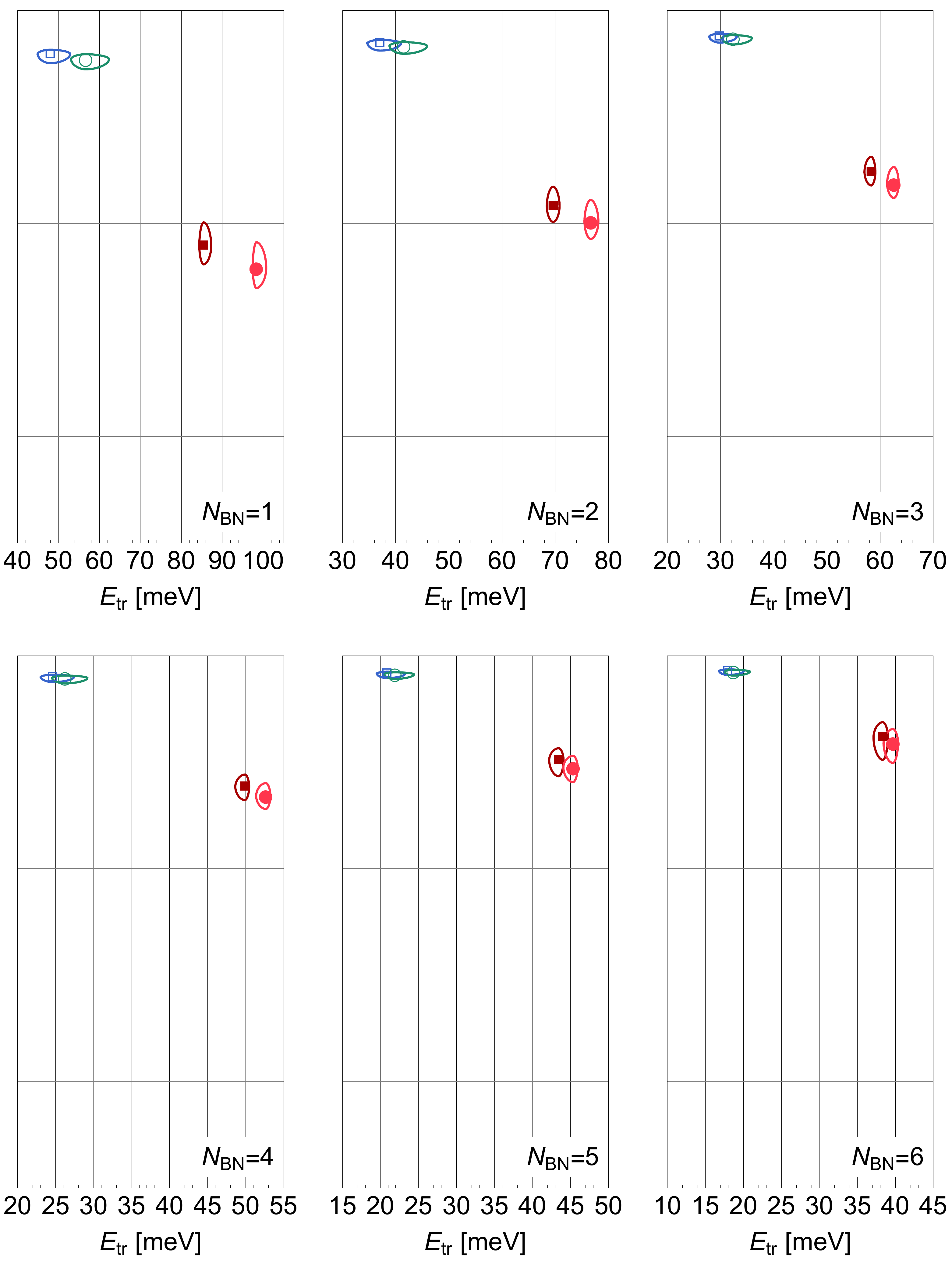}
	\caption{%
	  Relationship between $f_{0}^{j}$ and $E_{tr}^{j}$ for different interlayer separations characterized by the number of $h$-BN monolayers, $N_{\text{BN}}$.
	  The vertical range of the plots is $[0,1]$, so that the horizontal grid lines denote $f_{0}^{j} = 0.2,0.4,0.6,0.8$.
	}
	\label{fig:ind-foxy-vs-etr-rkc-1d}
  \end{figure}

  % Analyzing numerical results for f_0 for RK and C
  In Fig.~\ref{fig:ind-foxy-vs-etr-rkc-1d}, we present our calculations of $f_{0}^{j}$ for both the RK and Coulomb potentials for $N_{\text{BN}} = 1 - 6$.
  The $f_{0}^{j}$ are shown in separate plots for each value of $N_{\text{BN}}$.
  Our calculations show that increasing $N_{\text{BN}}$ leads to an increase in $f_{0}^{j}$ and a decrease in $E_{tr}^{j}$.
  Similar numerical studies of the indirect exciton in Xene~\cite{Brunetti2018b} and TMDC~\cite{Brunetti2018a} heterostructures with interlayer $h$-BN also indicated that $f_{0}^{j}$ is an increasing function of $N_{\text{BN}}$.
  % We note here, but will address in a following paragraph, that for $N_{\text{BN}} \geq 5$ we observe a peculiar broadening of the range of calculated values of $f_{0}^{x}$ and $E_{tr}^{x}$ as denoted by the widening red- and pink-colored ellipses.

  In general, $f_{0}^{y}$ does not change much as $N_{\text{BN}}$ increases because $f_{0}^{y}$ was already quite large for the direct exciton.
  On the other hand, since $f_{0}^{x}$ was small in the case of the direct exciton, we observe a significant increase in $f_{0}^{x}$ as $N_{\text{BN}}$ is incrementally increased.
  We also find that $E_{tr}[i,\text{C}, N_{\text{BN}}] > E_{tr}[i,\text{RK},N_{\text{BN}}]$, while $f_{0}^{j}[i,\text{RK},N_{\text{BN}}] > f_{0}^{j}[i,\text{C},N_{BN}]$, for any $i$ and $N_{\text{BN}}$.

  % REWRITE 7 & 8 ENTIRELY:
  Let us also mention an unusual trend in the range of calculated $f_{0}^{x}[i,p,N_{\text{BN}}]$ with respect to increasing $N_{\text{BN}}$ \textendash{} whereas the $f_{0}^{x}[i]$ become more tightly clustered as $N_{\text{BN}}$ increases from 1 to 5, the values become more spread out as $N_{\text{BN}}$ continues to increase from 5 to 8.
  The relative magnitudes of the $f_{0}^{x}[i]$ do not change with $N_{\text{BN}}$ nor with $p$ \textendash{} they are always related by $f_{0}^{x}[c] > f_{0}^{x}[d] > f_{0}^{x}[b] > f_{0}^{x}[a]$, as is the case with the direct exciton for $k = (\text{SS, HS, HE})$.
  Considering instead the incremental increase in $f_{0}^{x}[i]$ with $N_{\text{BN}}$ provides insight into the observed behavior.
  For example, the calculated values of $f_{0}^{x}[a]$ increase nearly linearly at small $N_{\text{BN}}$ before their growth is suddenly and strongly suppressed around $N_{\text{BN}} = 6$, i.e.\ $f_{0}^{x}[a,\text{RK},(1,2,3,4,5,6,7,8)] = (0.523,0.603,0.671,0.728,0.773,0.804,0.820,0.821)$.
  By contrast, $f_{0}^{x}[c]$ increases nearly linearly for all $N_{\text{BN}}$, i.e. $f_{0}^{x}[c,\text{RK},2] - f_{0}^{x}[c,\text{RK},1] = 0.067$, while $f_{0}^{x}[c,\text{RK},8] - f_{0}^{x}[c,\text{RK},7] = 0.057$.
  For comparison, the change in $f_{0}^{x}[b]$, which in general is only slightly larger than $f_{0}^{x}[a]$, starts to taper off between $N_{\text{BN}} = 7$ and $N_{\text{BN}} = 8$, suggesting that it is also approaching some kind of asymptotic limit, while $f_{0}^{x}[d]$, only slightly smaller than $f_{0}^{x}[c]$, also shows nearly linear growth throughout the range of $N_{\text{BN}}$ calculated here.

  We are therefore led to the conclusion that the incremental increase of $f_{0}^{x}[c]$ must be strongly suppressed for $N_{\text{BN}} \geq 8$, such that $f_{0}^{x}[c]$ approaches some constant value less than 1.
  On the other hand, it appears to be the case that the $f_{0}^{x}[a]$ has already converged towards its asymptotic value, which must be close to the calculated value of $0.821$ at $N_{\text{BN}} = 8$ \textendash{} meanwhile, $f_{0}^{x}[b]$ has an asymptotic maximum which is probably not much greater than $f_{0}^{x}[b,\text{RK},8] = 0.875$.
  Recalling $\mu_{c}^{x} > \mu_{d}^{x} > \mu_{b}^{x} > \mu_{a}^{x}$, it appears that the magnitude of $\mu^{j}$ is directly related to this asymptotic value of $f_{0}^{j}$ at large separations $N_{\text{BN}}$, and furthermore that the $N_{\text{BN}}$ at which this asymptotic value is reached also increases with $\mu^{j}$.

  Heterostructures of 2D materials exhibit a variety of interesting excitonic and optical behavior, including but not limited to the ability to tune the excitonic optical absorption strength and the corresponding transition energies.
  A more comprehensive study, one which, for example, systematically varies each input parameter individually over a broad yet physically plausible range, is necessary.
  By examining in detail how the excitonic properties change with respect to variations in the individual input parameters, we can deepen our understanding of which input parameters determine the maximal asymptotic value of $f_{0}^{j}$ and the interlayer separation $N_{\text{BN}}$ at which the asymptotic value is reached, and whether or not the asymptotic properties of $f_{0}^{j}$ can be further tuned by strain, dielectric environment, external electromagnetic fields, etc., and if so, by how much these quantities may change when these external tuning mechanisms are applied.

  % Analyzing %Differences for RK and C
  Let us now analyze in depth the effect that the choice of interaction potential has on the optical properties of the indirect exciton.

  Our calculations show that the percent differences for $E_{tr}^{j}$ and $f_{0}^{j}$ between the RK and C potentials generally decrease as $N_{\text{BN}}$ increases.
  First, we find that $E_{tr}^{j}[i,p_{\%},N_{BN}] > E_{n}[i,p_{\%},N_{\text{BN}}]$, for all $i,~j,~n$, and $N_{\text{BN}}$, i.e.\ the choice of interaction potential leads to a larger difference in the optical transition energies than in the corresponding individual eigenenergies, though in general the $E_{tr}^{j}[i,p,N_{\text{BN}}]$ follow the same trends with respect to increasing $N_{\text{BN}}$ as the eigenenergies themselves.
  These differences between the RK and Coulomb potentials decrease quickly with $N_{\text{BN}}$, from $E_{tr}^{j}[i,p_{\%},1] \approx (14.5\%,~16.5\%)$ for $j = (x,y)$, respectively, to $E_{tr}^{j}[i,p_{\%},5] \approx (4.2\%,~4.9\%)$.

  Turning now to the oscillator strengths, we observe some unusual deviations from the consistent patterns observed for $E_{tr}^{j}$.
  First, let us discuss the general relationship between $f_{0}^{j}$, $N_{\text{BN}}$, and $\mu_{i}$, returning later to the exceptions mentioned earlier.

  As mentioned earlier, since $f_{0}^{y}$ is already quite large for the direct exciton, it does not change significantly as $N_{\text{BN}}$ increases.
  By the same logic, the percent difference in $f_{0}^{y}[p_{\%}]$ is similarly small and decreases sharply as $N_{BN}$ increases.
  In particular, $f_{0}^{y}[p_{\%},(1,8)] \approx (1\%,~0.06\%)$, and furthermore, $f_{0}^{y}[c,p_{\%},N_{\text{BN}}] > f_{0}^{y}[b,p_{\%},N_{\text{BN}}] > f_{0}^{y}[a,p_{\%},N_{\text{BN}}] > f_{0}^{y}[d,p_{\%},N_{\text{BN}}]$, for all $N_{\text{BN}}$.
  By contrast, the relationship between $f_{0}^{x}[i]$ and the interaction potential is less straightforward.
  Whereas $f_{0}^{x}[(c,a),p_{\%},1] \approx (6.4\%,~8.87\%)$, corresponding to the minimum and maximum values, we find unexpectedly that the relationship is reversed at large interlayer separations, i.e.\ $f_{0}^{x}[(c,a),p_{\%},7] = (1.28\%,0.19\%)$.
  Furthermore, the percent difference $f_{0}^{x}[a,p_{\%}]$ actually increases from $N_{\text{BN}}=7$ to $N_{\text{BN}}=8$, from 0.19\% to 0.25\%.
  This is the only time that we observe an increase in the RK/C percent difference of any quantity with increasing $N_{\text{BN}}$.

  While these quantities may not be noteworthy on their own, they are analyzed in-depth here because of their unusual deviation from the trends which until now have consistently held true.
  It is unclear why only $f_{0}^{x}$ shows this abnormal progression, even as $E_{tr}^{x}$ and $f_{0}^{y}$ do not.

  \begin{table}[ht]
	\centering
	\caption{%
	  Dependence of the ratio $\widetilde{f}_{0}^{j}$, averaged over the $\mu_i$, on the number of $h$-BN monolayers, $N_{\text{BN}}$, for the RK and Coulomb potentials.
	  The values of $\widetilde{f}_{0}^{j}$ for each $\mu_i$ are tabulated in Table~\ref{tab:indf0muratiosfull} in Appendix~\ref{app:approach-scales}.
	  The units of $\widetilde{f}_{0}^{j}$ are [m$_{0}^{-1}$].
	}
	\begin{tabular}{r r c c c c c c}
	  \toprule[0.05em]\toprule[0.05em]
	  \multicolumn{2}{r}{{}}			& $N_{\text{BN}}=1$	& $N_{\text{BN}}=2$	& $N_{\text{BN}}=3$	& $N_{\text{BN}}=4$	& $N_{\text{BN}}=5$	& $N_{\text{BN}}=6$	\\ \midrule[0.05em]
	  \multirow{2}{*}{$\widetilde{f}_{0}^{x}[i_{avg.}]$}	& RK& 7.531				& 8.557				& 9.441				& 10.22				& 10.89				& 11.46				\\
	  {}											& C	& 6.963				& 8.155				& 9.136				& 9.981				& 10.71				& 11.33				 \\
	  \multirow{2}{*}{$\widetilde{f}_{0}^{y}[i_{avg.}]$}	& RK& 1.083				& 1.107				& 1.122				& 1.133				& 1.140				& 1.146				\\
	  {}											& C	& 1.072				& 1.101				& 1.119				& 1.131				& 1.139				& 1.145				 \\ \bottomrule[0.05em]
	  \bottomrule
	\end{tabular}
	\label{tab:indf0muratios}
  \end{table}

  In Table~\ref{tab:indf0muratios}, we present the calculated ratios $\widetilde{f}_{0}^{j}$, averaged over the four $\mu_i$, for all $N_{\text{BN}}$.
  We find that $\mathcal{A}^{x}[\text{RK}]$ increases from about 5.5\% to about 8.8\% as $N_{\text{BN}}$ increases from 1 to 8, while $\mathcal{A}^{y}[\text{RK}]$ increases from about 0.81\% to 0.87\% across the same range of $N_{\text{BN}}$.
  The results for the Coulomb potential are very similar, especially for $\mathcal{A}^{y}$ and at larger $N_{\text{BN}}$ for both $x$ and $y$, though we find that $\mathcal{A}^{x}[\text{C},1] \approx 5.1\%$, nearly a 10\% decrease compared to $\mathcal{A}^{x}[\text{RK},1]$.

  % Since $\alpha^{j}$ and $\mathcal{A}^{j}$ are obtained by multiplying a constant by a quantity which is proportional to $f_{0}^{j}$, the comparison between our results for the different $\mu_{i}$ and between the RK and Coulomb potentials is similar to the previous analysis on $f_{0}^{j}$.

\section{\label{sec:analysis}Analysis and discussion}

  % PARAGRAPH 1
  Now let us compare the properties of excitons in phosphorene to the properties of excitons in the TMDCs~\cite{Brunetti2018a} and the buckled 2D allotropes of silicon (Si), germanium (Ge), and tin (Sn), known as silicene, germanene, and stanene, and collectively as the Xenes~\cite{Brunetti2018b}.
  In Ref.~\onlinecite{Brunetti2018a}, the properties of indirect excitons in a TMDC/$h$-BN heterostructure (THT HS) were calculated using a similar method to the one used in this work.
  The study focused on four of the most common TMDCs, namely MoS$_{\text{2}}$, MoSe$_{\text{2}}$, WS$_{\text{2}}$, and WSe$_{\text{2}}$.
  Calculations were performed for a THT HS with $N_{\text{BN}} \in [1,9]$ using only the RK potential.
  The relevant material parameters were the exciton reduced mass, $\mu$, the 2D polarizability, $\chi_{2D}$, and the TMDC ML thickness, $l_{T}$.
  Many \textit{ab-initio} studies had previously calculated the material properties of the TMDCs, so for each material, two calculations were performed with two different values of the input parameters $\mu$ and $\chi_{2D}$.
  Out of the multitude of possible choices, the input parameters were chosen based on the combination of values which produced the largest and smallest exciton binding energy, corresponding to upper and lower bounds on all calculated quantities.
  In particular, the smallest reported value of $\mu$ and the largest reported value of $\chi_{2D}$ were used together to provide the lower bound, while the largest $\mu$ and smallest $\chi_{2D}$ found in the literature were used to provide the upper bound.

  % PARAGRAPH 2
  For indirect excitons in a THT HS, the binding energies were calculated to be between $E_{b}[\text{RK},N_{\text{BN}}=1] = (90-110,~100-105,~90-105,~90-110)$ meV in MoS$_{\text{2}}$, MoSe$_{\text{2}}$, WS$_{\text{2}}$, and WSe$_{\text{2}}$, respectively.
  Increasing the separation to $N_{\text{BN}} = 5$, the binding energies were reduced to between $50-70$ meV for all materials, decreasing to about $40-55$ meV at $N_{\text{BN}} = 8$.
  Comparing these values to the results shown in Fig.~\ref{fig:ind-en-vs-nbn-rkc}, we find that the binding energy of indirect excitons in a THT HS is smaller than in a PHP HS by about $5-10\%$.
  The $1s \to 2p$ optical transition energy of the indirect exciton in a THT HS was calculated to be about $E_{tr}[\text{RK},(1,5,8)] = (50-60,~30,~20)$ meV.
  By comparison, Fig.~\ref{fig:ind-foxy-vs-etr-rkc-1d} demonstrates that the anisotropic exciton reduced mass causes $E_{tr}^{x}$ ($E_{tr}^{y}$) to be significantly larger (smaller) than the analogous optical transition energy of the isotropic exciton.

  % PARAGRAPH 3
  To facilitate the comparison of the optical properties of excitons in different materials, we use the absorption factor $\mathcal{A}^{j}$ to control for the factor of $l$, which is different for each 2D material, in the denominator of Eq.~\eqref{eq:alphasimplify}.
  For a THT HS, the indirect exciton absorption factor was calculated to be $\mathcal{A}[\text{RK},1] = 2-3.7\%$, while in a PHP HS, we calculate $\mathcal{A}^{x}[p,1] \approx 5.1-5.5\%$ and $\mathcal{A}^{y}[p,1] \approx 0.81\%$, with the Coulomb potential yielding slightly smaller values of $\mathcal{A}$ than the RK potential.
  The calculated values of $\mathcal{A}$ in the TMDCs do not change significantly with increasing $N_{\text{BN}}$, reaching a maximum of about $2.5-4.4\%$ at $N_{\text{BN}} = 8$, because $f_{0}$ is already quite large at $N_{\text{BN}} = 1$, similar to the observed behavior of $f_{0}^{y}$ in a PHP HS\@.
  Again, we see here that the anisotropy of excitons in phosphorene leads to strongly enhanced (suppressed) optical activity under $x$- ($y$)-polarized excitations.

  % PARAGRAPH 4
  Turning now to the properties of excitons in Xenes, we note that a direct comparison is complicated by the uniquely tunable nature of excitons in the Xenes.
  Briefly, the buckled crystal structure of the Xenes allows the band gap, and therefore, the effective mass of charge carriers, to be tuned by an external electric field, $E_{\perp}$, oriented perpendicular to the plane of the Xene ML\@.
  As a result, the excitonic properties can be dramatically altered by changing the magnitude of the applied electric field.
  We will restrict the discussion here to a range of $E_{\perp}$ which lead to binding energies that are comparable to excitons in phosphorene.
  Also, a previous \textit{ab-initio} study predicted that the crystal structure of silicene became unstable around $E_{\perp} \approx 2.6$ V/\AA, so all calculations in Ref.~\onlinecite{Brunetti2018b} were performed for $E_{\perp} \leq 2.7$ V/\AA.

  % PARAGRAPH 5
  In Ref.~\onlinecite{Brunetti2018b}, the properties of both direct and indirect excitons were calculated.
  For direct excitons, results were obtained for freestanding (FS) Xene monolayers and for Si monolayers encapsulated by $h$-BN\@.
  The properties of indirect excitons were calculated using both the RK and Coulomb potentials in Xene/$h$-BN heterostructures, primarily focusing on silicene (SHS HS).

  % PARAGRAPH 6
  For the direct exciton in ML Xenes, it was calculated that $E_{b}[\text{Si},\text{FS}] \approx 740$ meV for $E_{\perp} \approx 1.5$ V/\AA, $E_{b}[\text{Ge},\text{FS}] \approx 740$ meV at $E_{\perp} = 2.7$ V/\AA, while the binding energy in FS Sn reached a maximum of about 550 meV.
  Compared to $E_{b}[\text{HE}] \approx 200$ meV in phosphorene, the direct exciton binding energy in HE Si reached a maximum of about 350 meV at the maximum electric field of $E_{\perp} = 2.7$ V/\AA, while the binding energy was about 200 meV for $E_{\perp} \approx 0.8-1.2$ V/\AA.

  Originally, calculations of $\alpha$ and $\mathcal{A}$ of excitons in the FS Xenes were performed for $\Gamma = 10^{13}$ s$^{-1}$, but for consistency we will instead assume $\Gamma = 10^{14}$ s$^{-1}$ as used here.
  Since the tuning mechanism of excitons in Xenes involves changing the charge carrier effective mass, the absorption coefficient and absorption factor are strongly suppressed at moderate to high electric fields, while the oscillator strength increases with increasing electric field.
  In general, when the electric field is large enough that the exciton binding energy is comparable to that of phosphorene, the value of $\mathcal{A}$ in the FS Xenes is only about 1\%, much weaker than $\mathcal{A}^{x}[\text{FS}]$ but comparable to $\mathcal{A}^{y}[\text{FS}]$.
  On the other hand, $\mathcal{A}[\text{Si},\text{HE}] \approx 2\%$ at $E_{\perp} = 1.0$ V/\AA, while $\mathcal{A}^{x}[\text{HE}] \approx 3\%$ and $\mathcal{A}^{y}[\text{HE}] \approx 0.6\%$, comparable to $\mathcal{A}$ in the FS Xenes.

  For indirect excitons in an SHS HS, the maximum binding energy at $E_{\perp} = 2.7$ V/\AA~was calculated to be about $E_b[\text{Si},p,1] \approx 150$ meV, not much bigger than the value of $E_b[i,p,1] \approx 120$ meV shown in Fig.~\ref{fig:ind-en-vs-nbn-rkc}.
  The data for $\mathcal{A}$ in an SHS HS again shows that indirect excitons are more optically active than direct excitons, where $\mathcal{A}[\text{Si},p,1] \approx 3-4\%$ for $E_{\perp} \approx 1$ V/\AA\@.
  Also, $\mathcal{A}[\text{Si},p,N_{\text{BN}}]$ was found to depend only weakly on the choice of interaction potential $p$, while the change in $\mathcal{A}$ with respect to $N_{\text{BN}}$ is again quite small in the SHS HS, comparable to $\mathcal{A}^{y}$.

  By comparing the properties of the anisotropic exciton in phosphorene to isotropic excitons in the TMDCs and Xenes, the effects of anisotropy are clearly emphasized.
  Whereas binding energies were mostly comparable in all three types of materials, the polarization-dependent optical properties of anisotropic excitons in phosphorene are drastically different from the optical properties of isotropic excitons in the TMDCs and Xenes.
  In particular, the small value of $\mu^{x}$ in phosphorene leads to a larger optical transition energy and significantly enhanced optical absorption, while the corresponding optical quantities for $y$-polarized excitations are much smaller than in isotropic excitons.

\section{\label{sec:conclusion}Conclusions}

  We study the optical properties of direct excitons in ML phosphorene, and of indirect excitons in a PHP HS, by calculating the $x$- or $y$-linear-polarization-dependent optical transition energies, oscillator strengths, absorption coefficients, and absorption factors.
  To calculate these properties, the eigenenergies and eigenfunctions of the exciton were calculated by solving the Schr\"{o}dinger equation using four different sets of anisotropic exciton reduced masses found in the literature.
  Additionally, we considered four different dielectric environments for the direct exciton corresponding to four common experimental (or theoretical, in the case of FS phosphorene) configurations.
  For the indirect exciton, the Schr\"{o}dinger equation was solved using both the Rytova-Keldysh and Coulomb interaction potentials, and at different interlayer separations $D$ corresponding to an integer number $N_{\text{BN}}=1-8$ of $h$-BN monolayers separating the ML phosphorene.
  Further analysis of our results for direct and indirect excitons was performed by examining how the results changed with respect to the change in exciton reduced mass, dielectric environment, choice of interaction potential, and interlayer separation.

  The intrinsic anisotropy of phosphorene manifests itself most noticeably in the optical properties of both direct and indirect excitons, where for direct excitons we predict that $\alpha^x > \alpha^y$ by as much as a factor of 8, with this difference decreasing to about a factor of four for ML phosphorene encapsulated by $h$-BN.
  By combining the calculated absorption coefficient with the known thickness of ML phosphorene, we predict that direct excitons in a single phosphorene ML may absorb as much as 3\% of incident $x$-polarized light, though this figure depends strongly on the 2D exciton concentration in the ML as well as on the line broadening of the excitonic transition.
  Analysis of the relationship between the absorption coefficient and the input parameters, and subsequent comparison to the optical properties of isotropic excitons in TMDCs and Xenes, suggests that the anisotropic mass is directly responsible for enhancing (suppressing) optical activity along the crystal axis with relatively light (heavy) exciton reduced mass.

  While exciton binding energies were comparable between the TMDCs, Xenes, and phosphorene, the excited states of the anisotropic exciton exhibit significant deviations from those of the isotropic exciton, where we find for the direct exciton that $E_{tr}^x[\text{FS}] > E_{tr}^y[\text{FS}]$ by nearly a factor of two, with this difference decreasing as dielectric screening increases.
  The exciton binding energy also strongly depends on the dielectric environment, where we calculate direct exciton binding energies of about 800 meV, 350 meV, and 200 meV, corresponding to FS phosphorene, uncapped phosphorene on an SiO$_2$ or $h$-BN substrate, and ML phosphorene encapsulated by $h$-BN.
  Furthermore, we find excellent agreement between our calculated binding energies and previous theoretical and experimental results.

  The increased spatial separation of the electron and hole in a PHP HS leads to a significant reduction in the indirect exciton binding energy compared to the direct exciton in the same dielectric environment.
  Specifically, we obtain an indirect exciton binding energy of about 120 meV in an PHP HS separated by only one ML of $h$-BN, compared to a direct exciton binding energy of 200 meV in HE phosphorene.
  Whereas the binding energy of the indirect exciton is reduced due to the increased interparticle separation, we find that the optical activity of the indirect exciton is enhanced compared to the direct exciton, and furthermore, that the oscillator strength is an increasing function of interlayer distance.
  As a result, we predict that indirect excitons in a PHP HS can absorb up to 5\% of an incident $x$-polarized excitation when separated by one ML of $h$-BN, increasing to more than 8\% absorption for 8 layers of $h$-BN, though we again note that the specific values of these quantities depend heavily on external factors such as exciton concentration and exciton broadening.

  In general, analysis of our results shows that increased dielectric screening leads to a decrease in all calculated quantities, i.e.\ the eigenenergies $E_n$, oscillator strength $f_0^j$, absorption coefficient $\alpha^j$, and absorption factor $\mathcal{A}^j$.
  The calculated binding energies are not particularly sensitive to the choice of $\mu_i$, but the optical properties can vary significantly depending on the relative magnitudes of $\mu^x$ and $\mu^y$.
  In particular, our results indicate that the optical transition energies $E_{tr}^{j}$, absorption coefficients $\alpha^{j}$, and absorption factors $\mathcal{A}^{j}$ are decreasing functions of the corresponding reduced mass $\mu^j$, while the oscillator strength $f_{0}^{j}$ is an increasing function of $\mu^{j}$.
  While the dependence of the optical properties on the $\mu^{j}$ is not completely straightforward, it is clear that any mechanism which affects the anisotropic charge carrier masses in phosphorene will in turn affect the optical properties of excitons in phosphorene.
  Considering that phosphorene is interesting to researchers precisely because of the external tunability of its properties via e.g.\ mechanical strain, an exhaustive study of the dependence of the excitonic and optical properties on parameters such as the anisotropic reduced mass, Rytova-Keldysh screening length, and environmental dielectric constant would be a welcome contribution to the literature.

  Our results represent the first comprehensive numerical calculations of the eigenenergies and optical properties of indirect excitons in a PHP HS with up to 8 layers of $h$-BN\@.
  Furthermore, our calculations support experimental observations and theoretical studies of the direct exciton binding energy in ML phosphorene.
  We then expand upon these results by analyzing the dependence of the optical properties of excitons in phosphorene on a variety of common input parameters.
  Our analysis indicates that the excitonic optical properties are highly sensitive to the anisotropic effective carrier masses, which can be tuned experimentally.
  Finally, our results demonstrate that an exhaustive study of the eigenstates and optical properties of the anisotropic exciton, in particular the dependence of these quantities on the input parameters shown in Table~\ref{tab:mutab}, is warranted.

  \section{Acknowledgements}

  The authors are grateful to acknowledge that this work is supported by the U.S. Department of Defense under Gran No. W911NF1810433.

\appendix

  \section{\label{app:approach-scales}Convenient simplifications to the analytical expressions for the excitonic optical quantities}

  In this Appendix we simplify the analytical expressions for the optical properties $\alpha^{j}$ and $\mathcal{A}^{j}$ presented in Sec.~\ref{sec:optics}.
  Examining Eqs.~\eqref{eq:f0xy},~\eqref{eq:alphasimplify}, and~\eqref{eq:afac}, we see that $f_{0}^{j}$ depends directly on the numerically calculated eigenenergies and eigenfunctions, while $\alpha^{j}$ and $\mathcal{A}^{j}$ are given by purely analytical expressions, provided $f_{0}^{j}$ is known.
  In other words, $f_{0}^{j}$ is the only optical quantity that depends directly on the numerical results \textendash{} on the other hand, $l$ and $\kappa$ are specified for a particular scenario, while $n_X$ and $\Gamma$ do not have specific values.
  As a result, $\alpha^j$ and $\mathcal{A}^j$ will exhibit the same qualitative behavior as the corresponding $f_{0}^{j}$ and $\mu_{i}^{j}$.
  By distinguishing $f_{0}^{j}$ and the associated $\mu^{j}$ from the constants and input parameters $l$, $\kappa$, $n_X$, and $\Gamma$, we aim to provide the reader with a simple way to calculate $\alpha^{j}$ and $\mathcal{A}^{j}$ using different parameters than those given in Table~\ref{tab:mutab}.
  These quantities, which we call the scale factors and denote by $C$ for the absorption coefficient $\alpha^{j}$, and $\mathcal{C}$ for the absorption factor $\mathcal{A}^{j}$, act as a sort of conversion factor between the cumbersome but straightforward analytical expressions for $\alpha^{j}$ and $\mathcal{A}^{j}$ and the values of e.g.\ $\mu_i$ and $f_{0}^{j}$ which are unique to our numerical results.

  Let us begin with Eq.~\eqref{eq:alphasimplify}, and as a first step separate the physical constants from the input parameters:

  \begin{align}
	\alpha^j & = C \left( \frac{n_X f_{0}^{j}}{\sqrt{\kappa} \mu^j l_{eff} \Gamma} \right),
	  \label{eq:alphac}\\
	C & = \frac{\pi e^2}{c m_0 \epsilon_0} = 3.335 \times 10^{-5}~\text{m}^{2}/\text{s},
	  \label{eq:calpha}
  \end{align}
  where $m_0$ is the rest mass of the electron.

  Now, the fraction within brackets in Eq.~\eqref{eq:alphac} contains all possible input parameters used in calculating $\alpha^{j}$, but we can further refine our expression for $C$ by recognizing that not every quantity shown in brackets in Eq.~\eqref{eq:alphac} is a free parameter.
  In particular, we consider only four values for $\kappa$ for the direct exciton and only one value for the indirect exciton as shown in Table~\ref{tab:mutab}.
  Similarly, we use only $l_{eff} = l_{phos}$ and $l_{eff} = 2 l_{phos}$ for the direct and indirect exciton, respectively.

  There are now five possible values of $C$ which are applicable to our results:

  \begin{align}
	C_{D} & = \frac{C}{\sqrt{\kappa} l_{phos}} = \begin{cases}
	  6.164 \times 10^{4}~\text{m/s},	& \text{FS} \\
	  3.980 \times 10^{4}~\text{m/s},	& \text{SS} \\
	  3.592 \times 10^{4}~\text{m/s},	& \text{HS} \\
	  2.788 \times 10^{4}~\text{m/s},	& \text{HE} \\
	\end{cases} \\
	  C_{I} & = \frac{C}{\sqrt{\kappa} (2 l_{phos}) } = 1.394 \times 10^{4}~\text{m/s},
	\label{eq:calphacases}
  \end{align}
  where the subscripts $D$ and $I$ denote direct and indirect excitons, respectively.

  Now Eq.~\eqref{eq:alphac} can be further simplified to:

  \begin{equation}
	\alpha_j = C_{D/I} \left( \frac{f_{0}^{j}}{\mu_{i}^{j}} \right) \left( \frac{n_X}{\Gamma} \right).
	\label{eq:alphacdi}
  \end{equation}

  Now, the foundation of our results, which consist of the numerically calculated eigenvalues and eigenfunctions, are effectively contained within the fraction $f_{0}^{j}/\mu_{i}^{j}$, for which we will use the notational shorthand $\widetilde{f}_{0}^{j} \equiv f_{0}^{j}/\mu_{i}^{j}$.
  On the other hand, $n_X$ and $\Gamma$ are essentially free parameters, insofar as the values given in Table~\ref{tab:mutab} are rough estimates meant to represent typical values of these quantities.
  Using the default values of $n_X$ and $\Gamma$ given in Table~\ref{tab:mutab}, we define the absorption coefficient scale factor as $\widetilde{C}_{D/I} = C_{D/I} \left( \frac{n_X}{\Gamma} \right)$ and obtain:

  \begin{align}
	\widetilde{C}_{D} & = \begin{cases}
	  3.082 \times 10^{6}~\text{m}^{-1},	& \text{FS} \\
	  1.990 \times 10^{6}~\text{m}^{-1},	& \text{SS} \\
	  1.796 \times 10^{6}~\text{m}^{-1},	& \text{HS} \\
	  1.393 \times 10^{7}~\text{m}^{-1},	& \text{HE} \\
	\end{cases} \nonumber \\
	  \widetilde{C}_{I} & = 6.969 \times 10^{6}~\text{m}^{-1}.
	\label{eq:ctildecalphacases}
  \end{align}

  We also note that $\widetilde{C}_{D/I}$ is independent of the $x$- or $y$-polarization of the excitation.
  Using Eqs.~\eqref{eq:calpha},~\eqref{eq:calphacases}, or~\eqref{eq:ctildecalphacases}, one can easily modify parameters such as $n_X$ or $\Gamma$ to match a particular scenario while still facilitating direct comparisons with the results presented in Sec.~\ref{sec:results}.

  Ultimately, the values of $\alpha^{j}$ presented in Sec.~\ref{sec:results} can therefore be calculated using the following expression:

  \begin{equation}
	\alpha^{j} = \widetilde{C}_{D/I} \left( \widetilde{f}_{0}^{j} \right).
	\label{eq:alphaeasycalc}
  \end{equation}

  The calculation of the absorption factor $\mathcal{A}^{j}$ can likewise be simplified:

  \begin{equation}
	\mathcal{A}^{j} = 1 - \exp \left[ - \widetilde{\mathcal{C}}^{D/I} \left( \widetilde{f}_{0}^{j} \right) \right],
	\label{eq:afacC}
  \end{equation}
  where the dimensionless quantity $\widetilde{\mathcal{C}}^{D/I} \equiv \widetilde{C}_{D/I} l_{eff}$ is given by,
  \begin{align}
	\widetilde{\mathcal{C}}^{D} & = \begin{cases}
	  1.668 \times 10^{-3},		&	\text{FS} \\
	  1.076 \times 10^{-3},		&	\text{SS} \\
	  9.717 \times 10^{-4},		&	\text{HS} \\
	  7.541 \times 10^{-3},		&	\text{HE} \\
	\end{cases} \nonumber \\
	  \widetilde{\mathcal{C}}^{I} & = 7.541 \times 10^{-3}.
	\label{eq:ctildeafaccases}
  \end{align}

  Finally we note that $\widetilde{\mathcal{C}}^{D/I} \ll \widetilde{f}_{0}^{j}$, so that the exponent in Eq.~\eqref{eq:afacC} is always much smaller than unity.
  Applying the well-known expansion of $e^{x}$ for small $x$, $e^{x} \approx 1 + x + (x^2)/2 + \dots$, the absorption factor can be approximated by:

  \begin{equation}
	\mathcal{A}^{j} \approx 1 - \left(1 + \left(- \widetilde{\mathcal{C}}^{D/I} \left( \widetilde{f}_{0}^{j} \right) \right) + \dots \right) = \widetilde{\mathcal{C}}^{D/I} \left( \widetilde{f}_{0}^{j} \right).
	\label{eq:afacapprox}
  \end{equation}
  This convenient approximation may prove useful for quickly estimating $\mathcal{A}^{j}$ from the values of $\widetilde{\mathcal{C}}^{D/I}$ given above, along with the values of $f_{0}^{j}$ and $\mu_{i}^{j}$ presented in Sec.~\ref{sec:results}.

  \begin{table}[ht]
	\centering
	\caption{%
	  Tabulated values of the ratio $\widetilde{f}_{0}^{j} \equiv f_{0}^{j}/\mu_{i}^{j}$ for each set of $\mu_i$, as well as the absorption coefficient scale factor, $\widetilde{C}_{D}$, from Eq.~\eqref{eq:ctildecalphacases}.
	  The units of $\widetilde{f}_{0}^{j}$ are [m$_{0}^{-1}$].
	}
	\begin{tabular}{c c c c c c c c c c}
	  \toprule[0.05em]\toprule[0.05em]
	  {}&$\widetilde{f}_{0}^{x}[a]$	&	$\widetilde{f}_{0}^{x}[b]$	&	$\widetilde{f}_{0}^{x}[c]$	&	$\widetilde{f}_{0}^{x}[d]$	&	$\widetilde{f}_{0}^{y}[a]$	&	$\widetilde{f}_{0}^{y}[b]$	&	$\widetilde{f}_{0}^{y}[c]$	&	$\widetilde{f}_{0}^{y}[d]$	& $\widetilde{C}_{D}$ [m$^{-1}]$ \\ \midrule[0.05em]
	  FS&9.96 & 9.56 & 7.51 & 8.21 & 0.93& 1.01& 1.33& 0.94 & $3.082 \times 10^{6}$\\
	  SS&7.20 & 7.00 & 5.81 & 6.18 & 0.88& 0.95& 1.24& 0.89	& $1.990 \times 10^{6}$\\
	  HS&6.36 & 6.20 & 5.25 & 5.53 & 0.86& 0.93& 1.21& 0.87 & $1.796 \times 10^{6}$\\
	  HE&4.31 & 4.24 & 3.77 & 3.86 & 0.78& 0.84& 1.08& 0.79	& $1.393 \times 10^{7}$\\ \bottomrule[0.05em]
	\end{tabular}
	\label{tab:dirf0muratiosfull}
  \end{table}

  \begin{table}[ht]
	\centering
	\caption{%
	  Tabulated values of the ratio $\widetilde{f}_{0}^{j}$ corresponding to each of the $\mu_i$, given in terms of the number of $h$-BN monolayers, $N_{\text{BN}}$, for the RK and Coulomb potentials.
	  The units of $\widetilde{f}_{0}^{j}$ are [m$_{0}^{-1}$].
	}
	\begin{tabular}{r c c c c c c c c c}
	  \toprule[0.05em]\toprule[0.05em]
	  \multicolumn{2}{c}{}				 &	$\widetilde{f}_{0}^{x}[a]$	&	$\widetilde{f}_{0}^{x}[b]$	&	$\widetilde{f}_{0}^{x}[c]$	&	$\widetilde{f}_{0}^{x}[d]$	&	$\widetilde{f}_{0}^{y}[a]$	&	$\widetilde{f}_{0}^{y}[b]$	&	$\widetilde{f}_{0}^{y}[c]$	&	$\widetilde{f}_{0}^{y}[d]$ \\ \midrule[0.05em]
	  \multirow{2}{*}{$N_{\text{BN}}=1$} & RK & 8.310 & 8.039 & 6.615 & 7.161 & 0.957 & 1.037 & 1.366 & 0.973 \\
	  {}								 & C  & 7.603 & 7.382 & 6.024 & 6.663 & 0.947 & 1.026 & 1.348 & 0.964 \\
	  \multirow{2}{*}{$N_{\text{BN}}=2$} & RK & 9.577 & 9.227 & 7.350 & 8.073 & 0.976 & 1.059 & 1.402 & 0.993 \\
	  {}								 & C  & 9.070 & 8.750 & 7.076 & 7.725 & 0.971 & 1.054 & 1.392 & 0.988 \\
	  \multirow{2}{*}{$N_{\text{BN}}=3$} & RK & 10.66 & 10.27 & 7.967 & 8.866 & 0.988 & 1.073 & 1.424 & 1.005 \\
	  {}								 & C  & 10.29 & 9.904 & 7.758 & 8.593 & 0.985 & 1.070 & 1.419 & 1.002 \\
	  \multirow{2}{*}{$N_{\text{BN}}=4$} & RK & 11.57 & 11.17 & 8.530 & 9.596 & 0.996 & 1.083 & 1.440 & 1.012 \\
	  {}								 & C  & 11.31 & 10.90 & 8.354 & 9.367 & 0.994 & 1.081 & 1.437 & 1.011 \\
	  \multirow{2}{*}{$N_{\text{BN}}=5$} & RK & 12.28 & 11.93 & 9.073 & 10.29 & 1.002 & 1.090 & 1.451 & 1.018 \\
	  {}								 & C  & 12.11 & 11.73 & 8.916 & 10.09 & 1.001 & 1.088 & 1.449 & 1.017 \\
	  \multirow{2}{*}{$N_{\text{BN}}=6$} & RK & 12.77 & 12.53 & 9.612 & 10.94 & 1.007 & 1.095 & 1.460 & 1.023 \\
	  {}								 & C  & 12.68 & 12.40 & 9.469 & 10.77 & 1.005 & 1.094 & 1.459 & 1.022 \\\bottomrule[0.05em]
	\end{tabular}
	\label{tab:indf0muratiosfull}
  \end{table}

\section{\label{app:eigenfunctions}Analysis of calculated eigenfunctions}

  Since the Schr\"{o}dinger equation~\eqref{eq:relschro} features anisotropy along the $x$ and $y$ axes, the calculated eigenfunctions are most conveniently characterized by the quantum numbers $n_x$ and $n_y$, in contrast to the isotropic case, where 2D polar coordinates are used and the eigenfunctions are described in terms of the principal and angular momentum quantum numbers, $n$ and $l$, analogous to the 2D hydrogen atom~\cite{Zaslow1967,Cotan2,Brunetti2018a}.
  The quantum numbers $n_x$ and $n_y$ corresponding to a particular eigenstate can be deduced by inspecting the eigenfunction and counting the number of times the eigenfunction changes sign along each axis.
  This is because eigenfunctions obey the empirical rule that the number of times the eigenfunction crosses $\psi(\mathbf{r}) = 0$ increases as the quantum number increases.

  When referring to the excitonic eigenstates in terms of the quantum numbers $n_x$ and $n_y$, we use the notation $(n_x,n_y)$, and denote the excitonic ground state by $(0,0)$.
  Similarly, the pairs $(1,0)$ and $(0,1)$ refer to the eigenstates where the exciton has absorbed one quantum of energy in the $x$- or $y$-directions, respectively \textendash{} we colloquially refer to these states as the ``first excited state in  ($x$ or $y$)''.

  Discussion of the eigenstates of the anisotropic exciton is further complicated by our computational method, which does not explicitly characterize the excitonic eigenstates in terms of the quantum numbers $n_x$ and $n_y$, or indeed, in terms of any set of quantum numbers.
  Instead, our calculations yield only the eigenvalues and eigenfunctions, sorted by decreasing eigenenergy.
  In the process of analyzing and discussing our results, it may be instructive to refer to a particular eigenstate not in terms of the quantum numbers $n_x$ and $n_y$, but by denoting it by its ``rank'' amongst all eigenstates produced by a particlar calculation.
  In this case we use the standard ket notation $\vert n \rangle$, that is, the ground state (which of course has the largest eigenenergy) is $\vert 1 \rangle$ and its eigenenergy is $E_{1}$, the first excited state (i.e.\ the state with the second-largest eigenenergy) is $\vert 2 \rangle$, with corresponding eigenenergy $E_{2}$, the second excited state (corresponding to the state with the third-largest eigenenergy, $E_{3}$) is $\vert 3 \rangle$, and so on.

  \begin{table}[ht]
	\centering
	\caption{%
	  Correspondence between the two notations for the calculated eigenstates: ranked in order of decreasing eigenenergy $\left( \vert n \rangle \right)$, and in terms of the quantum numbers $(n_x,n_y)$.
	  The correspondence between these notations is primarily determined by inspecting the calculated eigenfunctions, some of which are shown in Fig.~\ref{fig:efs-comp}.
	}
	\begin{tabular}{ccccccc}
	  \toprule[0.05em]\toprule[0.05em]
	  {}		&	$\vert 1 \rangle$	&	$\vert 2 \rangle$	&	$\vert 3 \rangle$	&	$\vert 4 \rangle$	&	$\vert 5 \rangle$	&	$\vert 6 \rangle$	\\ \midrule[0.05em]
	  $\mu_{a}$	&	$(0,0)$				&	$(0,1)$				&	$(0,2)$				&	$(0,3)$				&	$(0,4)$				&	$(1,0)$				\\
	  $\mu_{b}$	&	$(0,0)$				&	$(0,1)$				&	$(0,2)$				&	$(0,3)$				&	$(0,4)$				&	$(1,0)$				\\
	  $\mu_{c}$	&	$(0,0)$				&	$(0,1)$				&	$(0,2)$				&	$(1,0)$				&	$(0,3)$				&	$(0,4)$				\\
	  $\mu_{d}$	&	$(0,0)$				&	$(0,1)$				&	$(0,2)$				&	$(0,3)$				&	$(1,0)$				&	$(0,4)$				\\ \bottomrule[0.05em]
	\end{tabular}
	\label{tab:efsnxy}
  \end{table}

  \begin{figure}[ht]
	\centering
	\includegraphics[width=0.45\columnwidth]{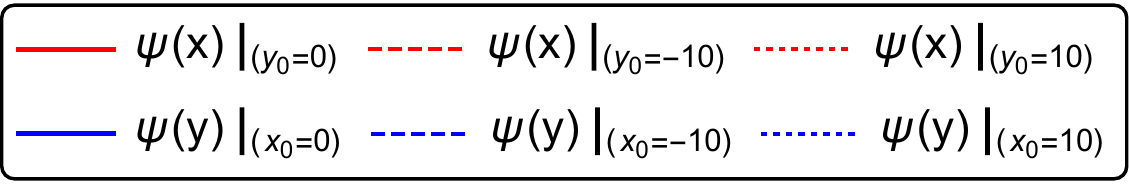}\\
	\includegraphics[width=0.95\columnwidth]{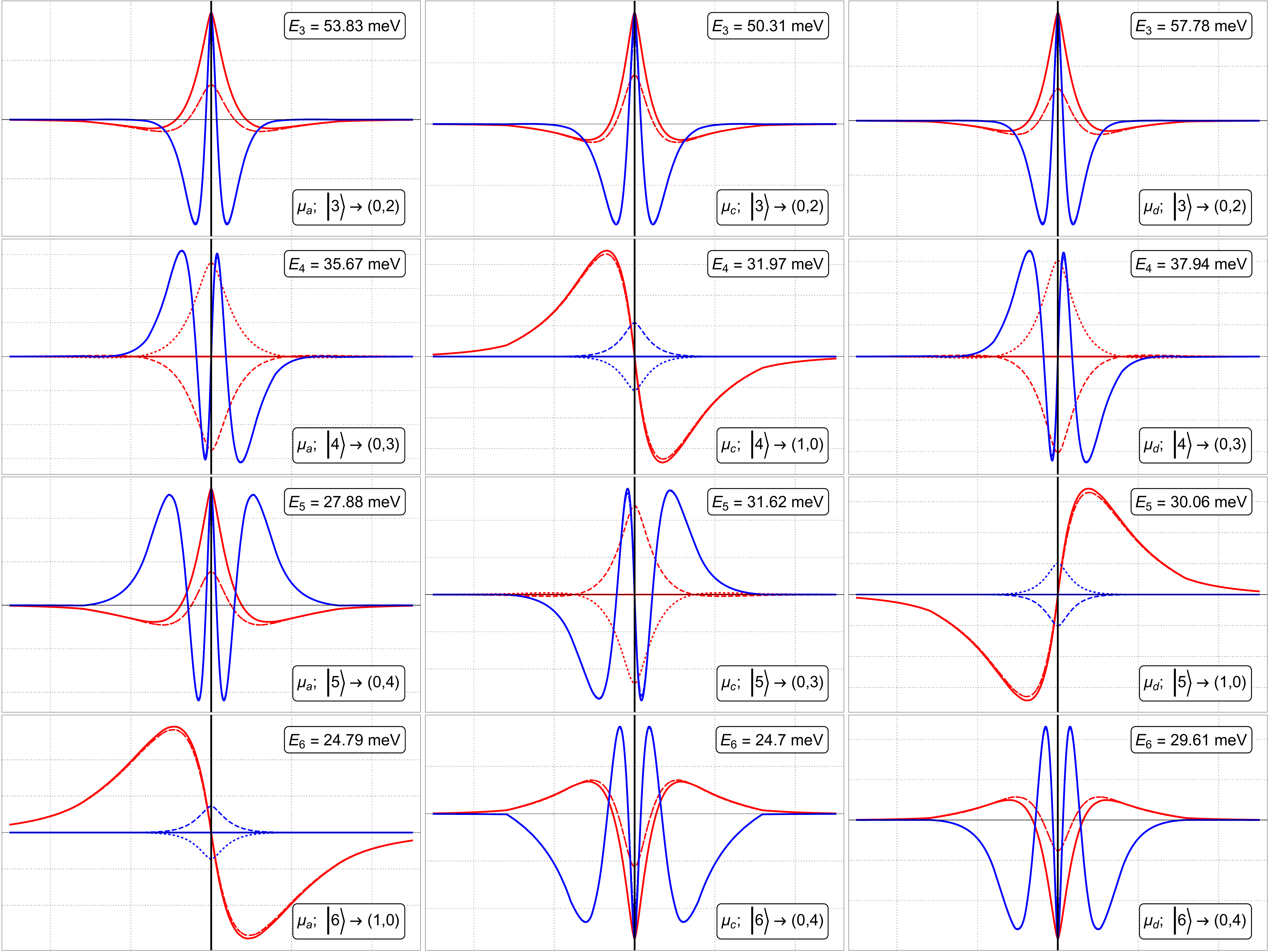}
	\caption{%
	  Comparison of the eigenfunctions for different eigenstates $\vert n \rangle$ and different choices of $\mu_i$.
	  The red lines represent plots along the $x$-axis with $y$ constant, while the blue lines are plotted along the $y$-axis with different constant values for $x$.
	}
	\label{fig:efs-comp}
  \end{figure}

  \begin{figure}[ht]
	\centering
	\includegraphics[width=0.51\columnwidth]{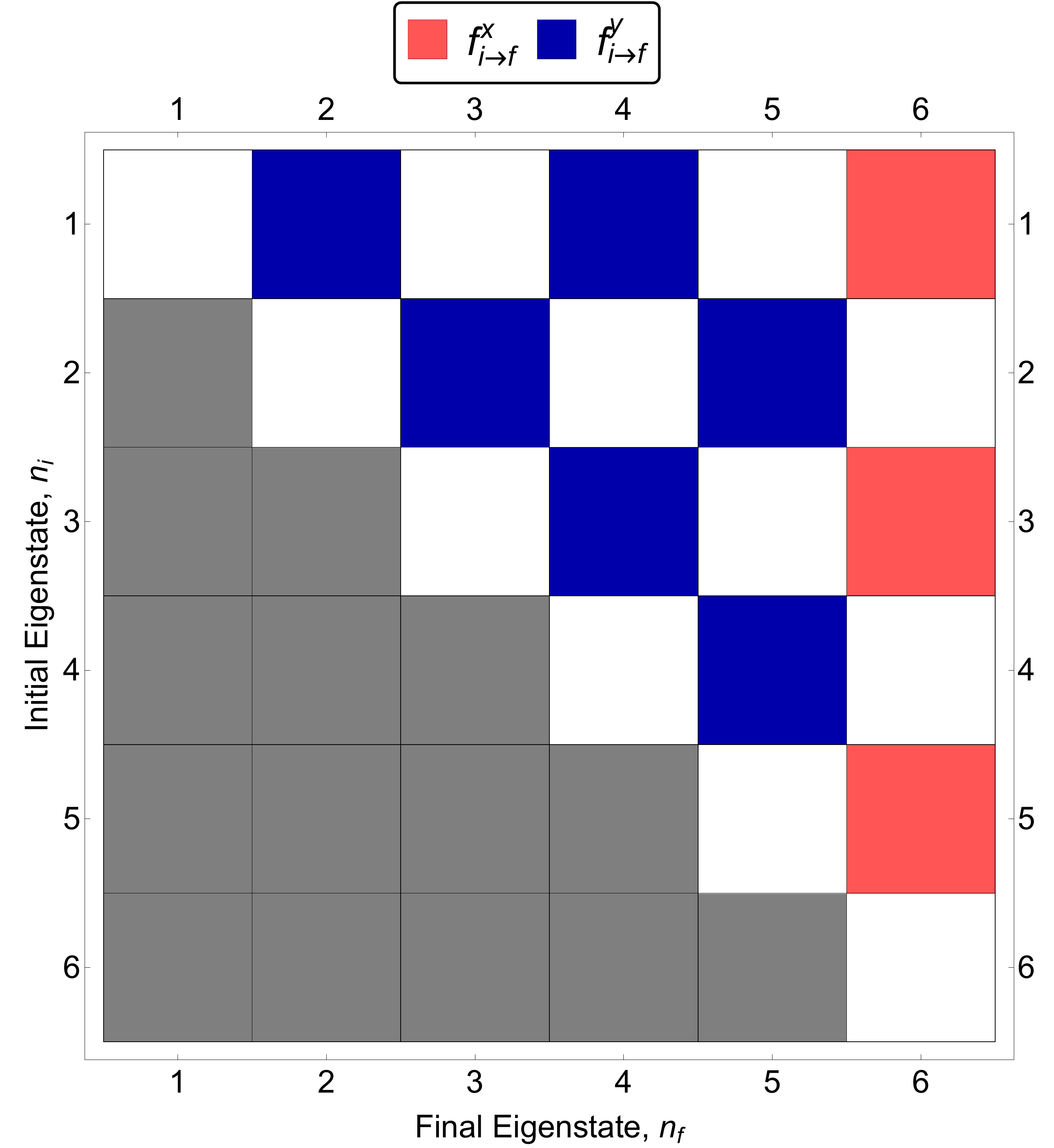}
	\caption{%
	  Allowed optical absorption transitions of the direct exciton using the set $\mu_a$ from any initial eigenstate $\vert n_i \rangle$ (corresponding to the rows) to a final eigenstate $\vert n_f > n_i \rangle$ (corresponding to the coulmns) given a linearly polarized excitation along either the $x$ or $y$ axes.
	  Allowed transitions are shown in red for $x$-polarized light and blue for $y$-polarized light.
	  Disallowed transitions are colored white.
	  Since the allowed optical transitions for radiative and absorptive processes are the same, the plot is symmetric along the diagonal.
	  Therefore, the bottom-left half of the plot has been colored gray to reduce visual clutter.
	}
	\label{fig:dir-allowed-trans-mu1}
  \end{figure}

  Curiously, we find that the ordering of the eigenstates with respect to the quantum numbers $n_x$ and $n_y$ is sensitive to the choice of $\mu_i$.
  In other words, the anisotropic reduced masses $\mu^x$ and $\mu^y$ change the eigenenergy of the first $x$ excited state $(1,0)$ relative to the eigenenergies of the higher excited states in $y$, in particular the states $(0,3)$ and $(0,4)$ as shown in Table~\ref{tab:efsnxy} and in Fig.~\ref{fig:efs-comp}.
  Further analysis of the eigenfunctions corresponding to $\vert n > 6 \rangle$ confirms that the ordering of the eigenstates in terms of $(n_x,n_y)$ differs depending on the relative magnitudes of $\mu^x$ and $\mu^y$.

  Finally, let us mention that while solutions to the Schr\"{o}dinger equation were obtained up to the $\vert 12 \rangle$ eigenstate, we have restricted our discussion and presentation of the results in the text to the first 6 eigenstates.
  Our reasons for this are fourfold: (i) to reduce visual clutter in the figures and emphasize the lower eigenstates which are more experimentally relevant; (ii) the optically active state $(1,0)$ can appear as late as $\vert 6 \rangle$, so we do not truncate our results before this state; (iii) the eigenenergies of states $\vert n > 6 \rangle$ change by a very small amount, and their optical activity is very strongly suppressed due to the presence of other allowed optical transitions to states with $\vert n \leq 6 \rangle$; (iv) eigenstates beyond approximately $\vert 8 \rangle$ strain our numerical methods and can occasionaly yield physically ambiguous or even non-sensical results.

  In Fig.~\ref{fig:efs-comp} we plot slices of the direct exciton eigenfunctions along the $x$- and $y$-axes.
  Also shown on the plots are the corresponding $\mu_i$, the eigenstate $\vert n \rangle$, and the eigenenergy of the state, $E_n$.
  We find that $\mu_{a}$ and $\mu_{b}$ share the same internal structure, and so plots for $\mu_{b}$ are not shown.
  The associations between $\vert n \rangle$ and $(n_x,n_y)$ shown in Table~\ref{tab:efsnxy} were determined by first examining the eigenfunctions (some of which are shown in Fig.~\ref{fig:efs-comp}) and counting the number of times the function changes sign along each axis, then cross-referencing those associations with the allowed and forbidden optical transitions of the anisotropic exciton, as determined theoretically in Ref.~\onlinecite{Rodin2014} and supported by our numerical results in Fig.~\ref{fig:dir-allowed-trans-mu1}.

  However, we note that the plots of the $(0,2)$ and $(0,4)$ states, corresponding to the states $\vert 3 \rangle$ (for all $\mu_i$) and either $\vert 5 \rangle$ (for $i=a,~b$) or $\vert 6 \rangle$ (for $i=c,~d$), respectively, clearly show that the eigenfunction changes sign twice with respect to the $x$ coordinate, suggesting that the states should have quantum number $n_x = 2$.
  Considering that these anomalous eigenstates appear before the $(1,0)$ eigenstate for all $\mu_i$, we conclude that this behavior is an aberration, and not to be interpreted as an appearance of a symmetric excited state in $x$ (e.g.\ a state characterized by $n_x = 2,4,6,\dots$).
  These eigenstates are optically dark, so it is difficult to assess how the abnormal behavior of the eigenfunction would affect calculations of the optical properties related to these states, if at all.

  In Fig.~\ref{fig:dir-allowed-trans-mu1}, the allowed and forbidden optical transitions between the first six eigenstates are shown in blue for $y$-polarized excitations and in red for $x$-polarized excitations.
  Counting from the top-left of the plot, the row numbers denote the initial eigenstate $\vert n_i \rangle$, while the column numbers correspond to the final eigenstate, $\vert n_f \rangle$.
  Boxes lying to the right (left) of the diagonal thus correspond to optical absorption (emission) transitions, where the location of each box in the array, specified by the ordered pair of (row,column) numbers, corresponds to the initial and final eigenstates of the transition.
  Each box corresponds to a possible optical transition, and the color of the box is based on the result of calculating $f_{0}^{x}$ and $f_{0}^{y}$.
  The box was colored red (blue) if $f_0^x$ ($f_0^y$) was calculated to be non-zero, and was colored white if neither calculation returned a non-zero result.

  Due to intrinsic error both in the numerical eigenfunctions themselves and resulting from numerical integration of the dipole transition matrix element, the oscillator strength was "non-zero" if it was greater than $10^{-4}$.
  The cutoff value of $10^{-4}$ was chosen after computing $f_0^j$ for all 12 calculated eigenstates and observing that the oscillator strengths of allowed transitions decreased by roughly an order of magnitude for each successive allowed transition from a given initial state.
  On the other hand, the numerical error in the calculated oscillator strengths for "dark" transitions was of the order of $10^{-10}$ or smaller for small $n_i$ and $n_f$, but reached as high as $10^{-7}$ for transitions involving eigenstates $n>8$.

  Qualitatively, the allowed and forbidden optical absorption transitions shown in Fig.~\ref{fig:dir-allowed-trans-mu1} agree exactly with the theoretically predicted optical selection rules of Ref.~\onlinecite{Rodin2014}.
  Apparently, the aforementioned anomalous eigenfunctions had no effect on the calculation of the optical selection rules using Eq.~\eqref{eq:f0xy}.

\bibliography{disser.bib}

\end{document}